\documentclass[11pt,a4paper]{article}
\usepackage{jheppub}
\usepackage{epstopdf}
\usepackage{multirow}
\usepackage{mathrsfs}
\usepackage{feynmp}
\usepackage{caption}          
\usepackage{subcaption}       

\usepackage[mathscr]{eucal}

\DeclareFontFamily{OT1}{pzc}{}
\DeclareFontShape{OT1}{pzc}{m}{it}{<-> s * [1.350] pzcmi7t}{}
\DeclareMathAlphabet{\mathpzc}{OT1}{pzc}{m}{it}

\def\sect#1{section~{\ref{#1}}}

\def\fig#1{fig.~{\ref{#1}}}

\def\app#1{appendix~{\ref{#1}}}

\def\tab#1{table~{\ref{#1}}}

\def\spa#1.#2{\left\langle#1\,#2\right\rangle}
\def\spb#1.#2{\left[#1\,#2\right]}

\def\bra#1{\langle #1 |}

\def\braket#1{\langle #1 \rangle}

\def\tree{{\rm tree}}

\def\Tr{\, {\rm Tr}}

\def\eps{\epsilon}

\def\nn{\nonumber}

\def\tf{\tilde{f}}

\def\eqn#1{eq.~\eqref{#1}}
\def\Eqn#1{Eq.~\eqref{#1}}
\def\eqns#1#2{eqs.~\eqref{#1} and~\eqref{#2}}
\def\Eqns#1#2{Eqs.~\eqref{#1} and~\eqref{#2}}

\def\be{\begin{equation}}
\def\ee{\end{equation}}
\def\bea{\begin{eqnarray}}
\def\eea{\end{eqnarray}}
\def\ba{\begin{eqnarray}}
\def\ea{\end{eqnarray}}
\def\beal{\begin{equation}\begin{aligned}}
\def\eeal{\end{aligned}\end{equation}}

\def\tree{{\rm tree}}

\newbox\charbox
\newbox\slabox
\def\s#1{{      
        \setbox\charbox=\hbox{$#1$}
        \setbox\slabox=\hbox{$/$}
        \dimen\charbox=\ht\slabox
        \advance\dimen\charbox by -\dp\slabox
        \advance\dimen\charbox by -\ht\charbox
        \advance\dimen\charbox by \dp\charbox
        \divide\dimen\charbox by 2
        \raise-\dimen\charbox\hbox to \wd\charbox{\hss/\hss}
        \llap{$#1$} }}

\def\nn{\nonumber}

\def\fmfsettings{
    \fmfset{thin}{1.2pt}
    \fmfset{arrow_len}{7pt}
    \fmfset{arrow_ang}{20}
    \fmfset{wiggly_len}{7pt}
    \fmfset{wiggly_slope}{70}
    \fmfset{curly_len}{6pt}
    \fmfset{dash_len}{8pt}
    \fmfset{dot_len}{3pt}
}

\title{Pure Gravities via  Color-Kinematics Duality for Fundamental Matter}
\author[a]{Henrik Johansson,}
\author[b]{Alexander Ochirov} 

\affiliation[a]{Theory Division, Physics Department, CERN, CH--1211 Geneva 23, Switzerland}
\affiliation[b]{Institut de Physique Th\'eorique, CEA--Saclay, F--91191 Gif-sur-Yvette cedex, France}

\emailAdd{henrik.johansson@cern.ch}
\emailAdd{alexander.ochirov@cea.fr}

\abstract{We give a prescription for the computation of loop-level scattering amplitudes
in pure Einstein gravity, and four-dimensional pure supergravities,
using the color-kinematics duality.
Amplitudes are constructed using double copies of pure (super-)Yang-Mills parts
and additional contributions from double copies of fundamental matter,
which are treated as ghosts.
The opposite-statistics states cancel the unwanted dilaton and axion in the bosonic theory, as well as the extra matter supermultiplets in the supergravity theories.
As a spinoff, we obtain a prescription for obtaining amplitudes
in supergravities with arbitrary non-self-interacting matter.
As a prerequisite, we extend the color-kinematics duality
from the adjoint to the fundamental representation of the gauge group.
We explain the numerator relations
that the fundamental kinematic Lie algebra should satisfy.
We give nontrivial evidence supporting
our construction using explicit tree and loop amplitudes,
as well as more general arguments.}

\preprint{CERN-PH-TH/2014-125 \\
\phantom{~} \hfill IPhT-t14/090}

\begin{document}
\maketitle

\pagebreak
\section{Introduction}
\label{introduction}

Recent progress in the understanding of scattering amplitudes
has revealed many novel mathematical structures
that are hidden from the standard Lagrangian approach.
One of the new structures,
discovered by Bern, Carrasco and one of the current authors~\cite{Bern:2008qj,Bern:2010ue},
is a duality between the kinematic and color content of gauge theories. Amplitudes can be constructed using kinematic numerators
that satisfy Lie-algebraic relations mirroring those of the color factors.
This has both interesting practical and conceptual consequences. It greatly simplifies the construction of loop amplitudes in (super-)Yang-Mills theory,
and more importantly, the duality relates multiloop integrands of (super-)Yang-Mills amplitudes
to those of (super-)gravity via a double-copy procedure. 
This has triggered a number of calculations and novel results
in ${\cal N}\ge 4$ supergravity that are otherwise extremely difficult to
obtain~\cite{Carrasco:2011mn,Bern:2012uf,Bern:2011rj,BoucherVeronneau:2011qv,Bern:2012cd,
Bern:2012uc,Bern:2012gh,Carrasco:2012ca,Carrasco:2013ypa,Bern:2013qca,Bern:2013uka}.

While the duality is a conjecture for loop amplitudes,
it is supported by strong evidence
through four loops in ${\cal N}=4$ SYM~\cite{Bern:2010ue,Bern:2012uf,Boels:2012ew}
and through two loops in pure Yang-Mills (YM) theory~\cite{Bern:2010ue,Bern:2013yya}.
Additional evidence comes from the attempts
to understand the kinematic algebra~\cite{Monteiro:2011pc,Boels:2013bi,Monteiro:2013rya}
and the Lagrangian formulation~\cite{Bern:2010yg,Tolotti:2013caa} of the duality.
The appearance of similar structures in string theory~\cite{BjerrumBohr:2009rd, Stieberger:2009hq,Tye:2010dd,Mafra:2011kj,Mafra:2011nw,Mafra:2012kh,Broedel:2013tta,
Ochirov:2013xba,Stieberger:2014hba, Mafra:2014oia},
Chern-Simons-matter theories~\cite{Bargheer:2012gv, Huang:2012wr,Huang:2013kca},
and, more recently, scattering equations~\cite{Cachazo:2012uq,Cachazo:2012da,Cachazo:2013gna,Cachazo:2013iea,Monteiro:2013rya}
sheds light on the universality of the duality.

The color-kinematics duality was originally formulated for gauge theories
with all fields in the adjoint representation of the gauge group.
The gravity theories that can be obtained
from the corresponding double copy of adjoint fields
are known as ``factorizable''~\cite{Carrasco:2012ca,Chiodaroli:2013upa}.
For ${\cal N}<4$ supersymmetry, such factorizable gravities have 
fixed matter content that goes beyond the minimal multiplets required by supersymmetry,
thus they are non-pure supergravities.
While this limitation is of little consequence at tree level,
for loop amplitudes the inability
to decouple or freely tune the extra matter content is a severe obstruction.
Many interesting gravity theories, such as Einstein gravity, pure ${\cal N}<4$ supergravity, or supergravities with generic matter,
have been inaccessible at loop level
from the point of view of color-kinematics duality. In contrast to this,
the loop-level construction of pure $4\le{\cal N}\le8$ supergravities
is well known~\cite{Bern:2010ue,Bern:2011rj,BoucherVeronneau:2011qv}.

Recently, the color-kinematics duality has been studied away from the case of purely-adjoint gauge theories. The duality has been fleshed out in the context of quiver gauge theories~\cite{Chiodaroli:2013upa} (see also refs.~\cite{Huang:2012wr,Huang:2013kca}), where fields are in the adjoint and bi-fundamental representations.
This construction of the color-kinematics duality
stays somewhat close to the adjoint case thanks to
the theory formulation through orbifold projections of adjoint parent theories.
The resulting double-copy prescription of ref.~\cite{Chiodaroli:2013upa}
gave many non-factorizable gravities, even if not pure supergravities.

In this paper, we eliminate the discussed limitations and
formulate a double-copy prescription for amplitude integrands in Einstein gravity,
pure supergravity theories, and supergravities with tunable
non-self-interacting matter.\footnote{By ``non-self-interacting matter'' we mean
no interactions beyond those required by general coordinate invariance
and supersymmetry; i.e. distinct matter supermultiplets do not interact directly with each other.} This is achieved by introducing the color-kinematics duality to gauge theories with matter fields in the fundamental representation.
We note that even at tree level, for four or more fundamental particles,
the double copy that follows from the fundamental color-kinematics duality is distinct
from the field-theory KLT relations~\cite{Kawai:1985xq,Bern:1998sv}.

Our main task is to obtain amplitudes for pure ${\cal N}<4$ supergravities,
including Einstein gravity. To do this, we combine double copies
of the adjoint and fundamental representations with a ghost prescription.
The double copies of fundamental matter are promoted to opposite-statistics states, which then cancel the unwanted matter content
in the factorizable gravities or adjoint double copies.
This construction critically relies on the double copies
of fermionic fundamental matter or supersymmetric extensions thereof.
As a naive alternative in the ${\cal N}=0$ case,
the fundamental scalar double copy works at one loop,
but starting at two loops it produces interactions
that no longer cancel the dilaton-axion states.
Although fundamental-scalar amplitudes appear to nontrivially satisfy
the color-kinematics duality, the corresponding double copies indicate
that the gravity amplitudes are corrected by four-scalar terms,
and possibly higher-order interactions,
which is consistent with the analysis of ref.~\cite{Johansson:2013nsa}.
In the present paper, we limit ourselves to the scalars
that are paired up with fermions within supersymmetric multiplets.

While we do not provide a rigorous proof of our framework,
we show its validity on various example calculations through two loops, and we give a multiloop argument using the effective R-symmetry of the tree amplitudes present in unitarity cuts.
As a warm-up, we discuss non-supersymmetric tree-level amplitudes.
At one loop, we obtain several duality-satisfying four-point numerators
with internal fundamental matter. Using these, we reproduce four-point one-loop ${\cal N}\le4$ supergravity amplitudes with external graviton multiplets, with or without matter in the loop.
As a highly nontrivial check at two loops,
we show in detail how our prescription cancels out the dilaton and axion
in the unitarity cuts of the four- and five-point two-loop amplitudes in Einstein gravity.

Interestingly, our approach directly generalizes to amplitudes in (super-)gravity theories
with arbitrary non-self-interacting matter: abelian vectors, fermions and scalars. Indeed,
once we obtain the tools to correctly subtract unwanted matter from loops,
we can reverse the procedure and instead add more matter,
introducing tunable parameters to count the number of matter states.
This generalization has a simple intuitive understanding:
the limited set of factorizable gravities can be attributed to the ``straightjacket''
imposed by the purely-adjoint color-kinematics duality with its unique color representation. As soon as we relax this, and ``complexify'' the duality to include the fundamental representation, we naturally gain access to a wider range of (super-)gravities.
Indeed, this observation is consistent with the quiver and orbifold constructions
of ref.~\cite{Chiodaroli:2013upa}.
That said, interesting gravitational matter amplitudes can also be obtained
through adjoint double copies~\cite{Bern:1999bx,Chiodaroli:2014xia}
that give an arbitrary amount of non-abelian vector multiplets.

This paper is organized as follows. We start \sect{colorkinematics} by the general definition of color-kinematics duality for the fundamental and adjoint representations, and proceed to illustrate its validity in four- and five-point non-supersymmetric tree amplitudes. In that part, we also introduce the relevant super-Yang-Mills (SYM) multiplets.
In \sect{FundDoubleCopy}, we give the prescription for the double-copy construction
of pure gravity theories, as well as the arguments in its favor.
In particular, in \sect{cutchecks},
we show explicitly how it removes the dilaton and axion
from the two-loop unitarity cuts in Einstein gravity. In \sect{multiloop}, we give a multiloop argument based on the effective R-symmetry of the tree amplitudes entering the unitarity cuts.
After these nontrivial checks,
in \sect{4ptYMsection}, we find all the color-dual four-point one-loop numerators for internal fundamental matter, and in \sect{gravity} we use these numerators to reproduce all one-loop four-point supergravity amplitudes with external graviton multiplets, with or without matter in the loop.

\section{Color-kinematics duality in the fundamental representation}
\label{colorkinematics}

In this section, we describe the general approach for exposing color-kinematics duality in scattering amplitudes of YM theories that have both adjoint and fundamental particles.
We start with a general formal setup, then give some concrete examples, and finish by explicitly listing the states of the YM theories under consideration.

\subsection{Kinematic numerators and color factors}

It is convenient to write the amplitudes of
$D$-dimensional (super-)Yang-Mills theory with fundamental matter as\footnote{We
use a different numerator normalization compared to ref.~\cite{Bern:2010ue}.
Relative to that work, we absorb one factor of $i$ into the numerator,
giving a uniform overall $i^{L-1}$ to the gauge and gravity amplitudes.}
\be
   {\cal A}^{\text{$L$-loop}}_m = i^{L-1} g^{m+2L-2}
      \sum_{i} \int\!\!\frac{d^{LD}\ell}{(2\pi)^{LD}}
                       \frac{1}{S_i} \frac{n_i c_i}{D_i} \,,
\label{BCJformYM}
\ee
where the sum runs over all $m$-point $L$-loop graphs with trivalent vertices of two kinds, (adj., adj., adj.) and (adj., fund., antifund.), corresponding to particle lines of two types: adjoint vector and fundamental matter.
For simplicity, each distinct assignment of vertices and internal lines is denoted by a unique graph label, here called $i$. The color factors $c_i$ are built out of products of structure constants $\tf^{abc}$ and generators $T^{a}_{j \bar k}$ matching the respective vertices.
Each denominator, $D_i$, is a product of the squared momenta of all internal lines of the graph, thus accounting for all the physical propagator poles.
$S_i$ are the usual symmetry factors that, for example, appear in Feynman diagrams.
The numerators $n_i$ are functions of momenta, polarizations and other relevant quantum numbers (excluding color).

\begin{figure}[t]
      \centering
      \includegraphics[scale=0.73]{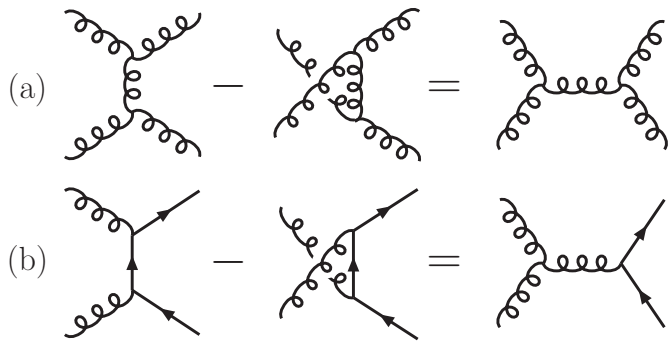}
      \vspace{-5pt}
\caption[a]{\small The kinematic algebra in the vector case~(a)
and in the fundamental matter case~(b).
Curly lines represent gluons or vector supermultiplets,
solid lines represents fermions, scalars or supersymmetric matter.
Alternatively, these diagrams describe standard Lie-algebra relations
for the color factors of the gauge group.}
\label{ChiralJacobiFigure}
\end{figure}

An important property of the representation \eqref{BCJformYM} is that the color factors satisfy linear relations of the schematic form
\be
   c_i-c_j=c_k \,,
\label{Jacobi}
\ee
which originate from the Jacobi identity and the commutation relation for the generators of the gauge group:
\begin{subequations} \begin{align}
      \tf^{dac} \tf^{cbe} - \tf^{dbc} \tf^{cae} & = \tf^{abc} \tf^{dce}
      \label{jacobi} \,, \\
      T^{a}_{i \bar \jmath} \, T^{b}_{j \bar k}   -
      T^{b}_{i \bar \jmath} \, T^{a}_{j \bar k} & = \tf^{abc} \, T^{c}_{i \bar k}
      \label{commutation} \,.
\end{align} \label{chiraljacobi} \end{subequations}
\!\!These relations are illustrated in \fig{ChiralJacobiFigure}.
Note that we work with the normalization convention in which the right-hand side of \eqn{commutation} is free of factors of $i$ and $\sqrt{2}$.
Additionally, we normalize the generators so that $\Tr(T^{a} T^{b})=\delta^{ab}$,
which together with \eqn{commutation} implies that
$\tf^{abc} = \Tr([T^{a},T^{b}]T^{c}) = \sqrt{2}i f^{abc}$,
where $f^{abc}$ are the structure constants more commonly found in the literature.
The above choices imply that the generators are hermitian,
$(T^{a}_{i \bar \jmath})^*=T^{a}_{j \bar \imath}$, and the structure constants are imaginary\footnote{Note that the choice of imaginary structure constants gives adjoint generators that are hermitian $(T^{a}_{\rm adj})_{bc} = (T^{a}_{\rm adj})_{cb}^*$,
where $(T^{a}_{\rm adj})_{bc} \equiv \tf^{bac}$.}
$(\tf^{abc})^*=-\tf^{abc}$.

The second property of the representation \eqref{BCJformYM} is the $Z_2$-freedom in the definition of the color factors:
the interchange of two adjoint indices of the same vertex amounts to a flip in the overall sign of the color factor, i.e.
\be
   c_j\rightarrow -c_j ~:~~~~~~  (\dots \tf^{abc} \ldots) \rightarrow
    - (\dots \tf^{abc} \ldots) = (\dots \tf^{bac} \ldots) \,.
\label{antisymmetry}
\ee

To streamline the notation, we adopt the convention that the interchange of a fundamental and an antifundamental leg also induces a sign flip of the color factors.
This amounts to introducing new generators $T^{a}_{\bar \imath j}$
that are trivially related to the standard ones:
\be
   T^{a}_{\bar \imath j} \equiv -T^{a}_{j \bar \imath} \,.
\label{antifundamentalgenerator}
\ee
When the generators have suppressed superscripts, we write them as 
$\overline{T}^{a} \equiv (T^{a})_{\bar \imath j}$.
With this convention we note that a color factor picks up a minus sign
whenever two legs attached to a cubic vertex of any representation are interchanged.
In practice, this means that we can identify the relative sign of a vertex with the cyclic orientation of its three legs, thus clockwise and counterclockwise orderings have different overall signs.

While we mainly consider matter particles to be in the fundamental representation of the gauge group, it is sometimes convenient to work with adjoint matter. The latter is obtained by simply swapping the generators as follows:
\be
   T^{a}_{i \bar \jmath} \rightarrow \tf^{bac}
   ~~{\rm with}~~i\rightarrow b,\, \bar \jmath \rightarrow c \,.
\label{representationswap}
\ee
By applying this rule, one goes from a complex representation to a real one,
which makes the color factors of matter and antimatter fields indistinguishable from one another.
Let us illustrate how this affects a one-loop amplitude in the form~\eqref{BCJformYM},
as is relevant for the explicit calculations of \sect{4ptYMsection}.

\begin{figure}[t]
      \centering
      \includegraphics[scale=0.75]{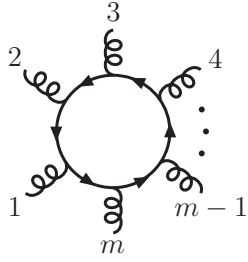}
      \vspace{-5pt}
\caption[a]{\small A typical one-loop graph with external adjoint and internal fundamental particles.}
\label{NgonParentFundFigure}
\end{figure}

Consider the one-loop ``ring diagram'' shown in \fig{NgonParentFundFigure}.
Its color factor is
\be
   c_{12 \dots m} = \Tr(T^{a_1}T^{a_2} \dots T^{a_m})
   \equiv T^{a_1}_{i \bar \jmath}T^{a_2}_{j \bar k} \dots T^{a_m}_{l \bar \imath} \,.
\label{ringdiagram}
\ee
Now consider the same diagram, but with the internal arrows reversed.
Throughout this paper, we will denote the operation of reversing fundamental matter arrows (matter $\leftrightarrow$ antimatter) by a bar, hence the corresponding color factor is given by
\be
   \overline{c}_{12 \dots m} = \Tr(\overline{T}^{a_1}\overline{T}^{a_2} \dots \overline{T}^{a_m})
   \equiv T^{a_1}_{\bar \imath j}T^{a_2}_{\bar j  k} \dots T^{a_m}_{\bar l i}
                            = (-1)^m \, \Tr (T^{a_m} \dots T^{a_2} T^{a_1}) \,.
\label{ringdiagramconj}
\ee
These two diagrams have the same propagators but different color factors.
However, if we promote them to the adjoint representation using the rule \eqref{representationswap},
both color factors \eqref{ringdiagram} and \eqref{ringdiagramconj} are mapped to the same object
\be
   c^{\rm adj}_{12\dots m}=\tf^{b a_1 c}\tf^{c a_2 d}\dots \tf^{e a_m b} \,.
\label{ringdiagramadjoint}
\ee
Since the two graphs have now identical color factors,
it is natural to repackage them as follows:
\be
      c_{12\dots m}n_{12\dots m} + \overline{c}_{12\dots m} \overline{n}_{12\dots m}
      ~~\longrightarrow ~~
      c^{\rm adj}_{12\dots m} (n_{12\dots m} + \overline{n}_{12\dots m})\equiv c^{\rm adj}_{12\dots m} n^{\rm adj}_{12\dots m} \,,
\label{adjointPromotion}
\ee
where $\overline{n}_{12\dots m}$ denotes the antimatter contribution, and $n^{\rm adj}_{12\dots m}$ defines an effective numerator for adjoint matter inside the loop.
As can be easily verified, for generic one-loop numerators with adjoint external states and internal matter, the same convenient relation exists between the fundamental and adjoint contributions,
\be
      n_i^{\rm adj} = n_i + \overline{n}_i \,.
\label{nplusnbar}
\ee
In the case that the matter multiplet is effectively non-chiral, implying that $n_i$ is CPT-invariant and $n_i=\overline{n}_i$ from the start, \eqn{adjointPromotion} gives a numerator from a non-minimal adjoint-matter multiplet.
Then the definition $n_i^{\rm adj} = n_i= (n_i+\overline{n}_i)/2$ may be more convenient to use, since it gives the minimal-multiplet contribution.

Returning to the general multiloop amplitude, we note that thus far the formula~\eqref{BCJformYM} is a trivial rewrite of standard (super-)Yang-Mills perturbation theory.
The only minor change is that we have implicitly absorbed the quartic interactions into the cubic graphs.
To expose a duality between color and kinematics~\cite{Bern:2008qj,Bern:2010ue},
we need to enforce nontrivial constraints on the kinematic numerators,
effectively making them behave as objects of a kinematic Lie algebra.
In particular, we demand that the numerator factors obey
the algebraic relations that are ubiquitous for color factors of any  Lie algebra.
Namely, the Jacobi/commutation relations~\eqref{Jacobi}
and the antisymmetry~\eqref{antisymmetry} under the interchange of legs attached to a single vertex.  These imply dual relations schematically written as follows:
\begin{subequations} \begin{align}
      n_i - n_j = n_k ~~~&\Leftrightarrow~~~ c_i - c_j=c_k \,, \\
      n_i \rightarrow -n_i  ~~~&\Leftrightarrow~~~ c_i \rightarrow -c_i \,.
\end{align} \label{duality}%
\end{subequations}
For every such three-term color identity,
there is a corresponding kinematic numerator identity,
and for each sign flip in a color factor,
there is a corresponding sign flip in the kinematic numerator.

Note that the three-term identities for numerators take the same pictorial form as the Lie-algebra identities for color factors, as shown in \fig{ChiralJacobiFigure}. However, they now represent the kinematic constraints that can be consistently imposed on the interactions involving not only adjoint but also fundamental particles. That these constraints are consistent with the amplitudes of various (super-)Yang-Mills theories is highly nontrivial. In the purely-adjoint case, the constraints lead to so-called BCJ relations between tree-level amplitudes, which have been proven in the context of string~\cite{BjerrumBohr:2009rd,Stieberger:2009hq} and field theory~\cite{Feng:2010my,Chen:2011jxa,Du:2011js, Cachazo:2012uq}.

\begin{figure}[t]
      \centering
      \includegraphics[scale=0.73]{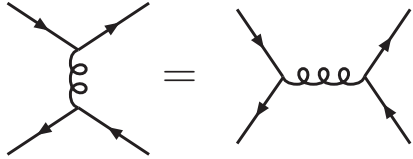}
      \vspace{-5pt}
\caption[a]{\small An additional two-term kinematic identity
that can be imposed on the numerators
for indistinguishable fermions in spacetime dimensions $D=3,4,6,10$,
or for indistinguishable complex scalars in any dimension.
However, this relation is not mandatory
for the color-kinematics duality.
In fact, the corresponding color factors will in general not satisfy this relation.}
\label{TwoTermIdFigure}
\end{figure}

In addition to the three-term Jacobi/commutation identities, we find that we can impose a nontrivial two-term constraint, shown in \fig{TwoTermIdFigure}, valid for fundamental matter numerators in particular situations. For the four-point pure-matter amplitude, corresponding to the diagrams in \fig{TwoTermIdFigure}, the identity is simply
\be
   n_t = n_s\,,
\label{4ptTwoTerm}
\ee
where $n_t$ and $n_s$ are the $t$- and $s$-channel kinematic numerators, respectively. 
For indistinguishable complex scalars, whose Lagrangian includes both the minimal coupling to the gauge field and a quartic-scalar term (see \app{AppendixA}), the two-term identity holds in any spacetime dimension. It is also applicable for indistinguishable minimal-Lorentz-representation fermions in $D=3,4,6,10$ dimensions. This can be shown by considering the real or non-chiral version of the two-term identity, i.e. by converting between complex and real scalars, or Weyl and Majorana fermions.  For a real scalar, the identity becomes equivalent to the adjoint Jacobi relation for kinematic numerators, which holds at tree level because the (YM + scalar)-theory is equivalent to pure YM theory in the higher-by-one spacetime dimension. For the non-chiral (Majorana) fermion, the identity again becomes equivalent to the adjoint Jacobi relation for kinematic numerators, which holds in $D=3,4,6,10$ due to a Fierz identity, as shown in ref.~\cite{Chiodaroli:2013upa}, where the numerators and amplitudes belong to ${\cal N}=1$ SYM theory.

Although this identity can be consistently imposed in, say, $D=4$, it is actually not necessary for obtaining an amplitude that satisfies the color-kinematics duality. Without going into details, this can be seen from the observation that the corresponding color factors will in general not satisfy this relation.
In terms of the generators, it would correspond to
\be
   T^a_{i \bar \jmath} T^a_{k \bar l} \stackrel{?}{=}
   T^a_{i \bar l} T^a_{k \bar \jmath} \,.
\label{ColorTwoTerm}
\ee
Although fundamental representations of generic gauge groups do not obey this relation, the generator of U(1) does, as well as generators of particular tensor representations of U($N$).\footnote{For example,
consider the symmetric (antisymmetric) part
of the $N\otimes N$ representation of U($N$) with generators 
$\widetilde{T}^{a}_{\beta \bar \gamma} = \Tr(M_{\beta} T^{a} M_{\gamma}^{*})$,
where $M_{\beta}$ form a basis of symmetric (antisymmetric) $N$-by-$N$ matrices.}
A sufficient condition for exhibiting color-kinematics duality is that
the kinematic numerators obey the same {\it general} relations as the color factors.
Since \eqn{ColorTwoTerm} is not a general identity,
imposing it on the numerators is optional.
Indeed, for fermion amplitudes in general spacetime dimension,
the rule~\eqref{4ptTwoTerm} would be inconsistent.
For the purpose of dimensional regularization, where $D=4-2\epsilon$,
we also do not want to rely on a four-dimensional identity.

An important feature of the amplitude representation~\eqref{BCJformYM} is its non-uniqueness. As the numerators of individual diagrams are not gauge-invariant, they have a shift freedom that leaves the amplitude invariant. It is known as generalized gauge invariance~\cite{Bern:2008qj, Bern:2010ue}. More precisely, a shift of the numerators $n_i \rightarrow n_i + \Delta_i$
does not change the amplitude, provided that $\Delta_i$ satisfy
\be
   \sum_{i} \int\!\!\frac{d^{LD}\ell}{(2\pi)^{DL}}
                    \frac{1}{S_i} \frac{ \Delta_i c_i }{D_i} =0 \,.
\label{YMshift}
\ee
The duality conditions~\eqref{duality} constrain to some extent the freedom of the numerators, but not entirely.
The remaining freedom to move terms between different diagrams means that there usually exist many different amplitude representations that obey color-kinematics duality. This can be useful for finding representations that make various properties manifest. But more importantly, generalized gauge invariance provides a guidance for constructing gravity amplitudes out of the gauge-theory numerators.  Any prescription for this must preserve the generalized gauge invariance: the gravity amplitudes cannot depend on the arbitrariness of the gauge-theory numerators. We will use this fact in \sect{GaugeInvarianceSect}.

In general, we know that the color-kinematics duality provides a simple method~\cite{Bern:2008qj,Bern:2010ue} for constructing gravity amplitudes via a double copy of the numerators:
\be
   {\cal M}^{\text{$L$-loop}}_m
      = i^{L-1} \Big(\frac{\kappa}{2}\Big)^{m+2L-2}
        \sum_{i} \int\!\!\frac{d^{LD}\ell}{(2\pi)^{DL}}
        \frac{1}{S_i} \frac{n_i n_i'}{D_i} \,,
\label{BCJformGravityAdj}
\ee
where the gravity theory is determined by the choice of a pair of two gauge theories, encoded in the numerators $n_i$ and $n_i'$. 
An alternative way~\cite{Bern:2011rj} to think of this prescription is to take the gauge-theory amplitude~\eqref{BCJformYM} and replace $c_i \rightarrow n_i'$.
For this to be consistent, $n_i'$ must satisfy the same general algebraic relations as $c_i$, i.e. the numerators must obey color-kinematics duality. However, it is known that the $n_i$ in \eqn{BCJformGravityAdj} need not satisfy the duality~\cite{Bern:2010ue,Bern:2010yg}. In \sect{FundDoubleCopy}, we will present a generalization of \eqn{BCJformGravityAdj} specific to pure gravity theories, and also explain how one can generalize the replacement $c_i \rightarrow n_i'$ to get gravity amplitudes with generic non-self-interacting matter.

\subsection{Some concrete examples}
\label{treeexamples}

In order to make the discussion in the previous section more concrete, and to connect to the standard Feynman diagram expansion, here we give examples of color-kinematics duality for two simple tree-level amplitudes.

\subsubsection{Four-point tree amplitudes}
\label{tree4pt}

Consider the four-quark tree-level amplitude in QCD, $q \bar{q}q \bar{q}$, where we set the quark masses to zero. If we take legs 1 and 3 to be quarks and legs 2 and 4 to be antiquarks and do not distinguish between flavors, then the amplitude is the sum of the following two Feynman diagrams:
\be
      \parbox{64pt}{
      \begin{fmffile}{qqQQ1} \fmfframe(12,12)(-12,12){
      \fmfsettings
      \begin{fmfgraph*}(50,40)
            \fmflabel{$1^-\!, i\!\!\!\!\!$}{q1}
            \fmflabel{$2^+\!, \bar \jmath\!\!\!\!\!$}{q2}
            \fmflabel{$\!\!\!\!\!\!\!\!3^-\!, k$}{q3}
            \fmflabel{$\!\!\!\!\!\!\!\!4^+\!, \bar l$}{q4}
            \fmfleft{q1,q2}
            \fmfright{q4,q3}
            \fmf{plain_arrow}{q2,v1,q1}
            \fmf{plain_arrow}{q4,v3,q3}
            \fmf{curly,tension=0.3}{v3,v1}
      \end{fmfgraph*} }
      \end{fmffile}
      }
      = -i\,T_{i \bar \jmath}^a T_{k \bar l}^a \,
         \frac{\braket{13} [24]}{s} = -i \frac{c_s n_s}{s} \,, ~
      \parbox{60pt}{
      \begin{fmffile}{qqQQ2} \fmfframe(12,12)(-20,12){
      \fmfsettings
      \begin{fmfgraph*}(50,40)
            \fmflabel{$1^-\!, i\!\!\!\!\!$}{q1}
            \fmflabel{$2^+\!, \bar \jmath\!\!\!\!\!$}{q2}
            \fmflabel{$\!\!\!\!\!\!\!\!3^-\!, k$}{q3}
            \fmflabel{$\!\!\!\!\!\!\!\!4^+\!, \bar l$}{q4}
            \fmfleft{q1,q2}
            \fmfright{q4,q3}
            \fmf{plain_arrow}{q4,v1,q1}
            \fmf{plain_arrow}{q2,v2,q3}
            \fmf{curly,tension=0.3}{v2,v1}
      \end{fmfgraph*} }
      \end{fmffile}
      }
      = -i\,T_{i \bar l}^a T_{k \bar \jmath}^a \,
         \frac{\braket{13} [24]}{t} = -i \frac{c_t n_t}{t} \,.
\label{qqQQ}
\ee
Here and below, we use the standard notation for spinor products and the Mandelstam invariants:
\be
 s_{ij} \equiv (k_i+k_j)^2= \braket{ij} [ji] \,, \quad
      s \equiv (k_1+k_2)^2 \,, \quad
      t \equiv (k_2+k_3)^2 \,, \quad
      u \equiv (k_1+k_3)^2 \,.
\label{Mandelstam}
\ee

An important property of the above amplitude is that the corresponding $u$-channel diagram cannot be constructed from Feynman rules. It is kinematically zero because of the external helicity choices, and similarly forbidden because the color state of the $u$-channel state has no overlap with the adjoint representation. This meshes well with the fact that these diagrams comprise a representation where kinematics is dual to color. That the Feynman rules land exactly on such a representation is a somewhat accidental property of the particularly simple amplitude, and it can be traced back to the gauge invariance of the individual diagrams. Indeed, there exist no (generalized) gauge freedom that can shuffle terms between the two numerators. This is clear if we consider the quarks to have two different flavors,  e.g. $q \bar{q} q' \bar{q}'$. Then only the first of the above diagrams is allowed in the amplitude and thus must be gauge invariant by itself.

Let us take a step back and carefully check that we have an amplitude representation of the form described in the previous section. First of all, as the only adjoint particle is the intermediate gluon, neither of the two identities of the kinematical algebra (shown in \fig{ChiralJacobiFigure}) is applicable. However, we do have the option to enforce the two-term identity in $D=4$. But since the numerators are already unique, the identity should be automatically true. Indeed, looking at the two numerators we have
\be
   n_s = \braket{13} [24] = n_t \,.
\ee

This example may seem trivial on the YM side,
but it is highly nontrivial that one can now construct gravity amplitudes
using these numerators.
For example, the four-photon amplitude\footnote{We use plain $M$
to denote gravity amplitudes with omitted factors of $\kappa/2$, i.e. $\kappa=2$.}
is given by
\be
   M_4^{\rm tree}(1^-_{\gamma},2^+_{\gamma},3^-_{\gamma},4^+_{\gamma})
      = -i \frac{n_s^2}{s} - i \frac{n_t^2}{t}
      = -i \braket{13}^2 [24]^2 \Big(\frac{1}{s}+\frac{1}{t}\Big) \,.
\ee 
And if we have two types of U(1)-vectors, say $\gamma$ and $\gamma'$, we have the following amplitude
\be
   M_4^{\rm tree}(1^-_{\gamma},2^+_{\gamma},3^-_{\gamma'},4^+_{\gamma'})
      = -i \frac{n_s^2}{s} = -i \braket{13}^2 [24]^2 \frac{1}{s} \,,
\label{TwoPhotonexample}
\ee
simply obtained by removing the $t$-channel graph.

\begin{figure}[t]
      \centering
      \includegraphics[scale=1.00]{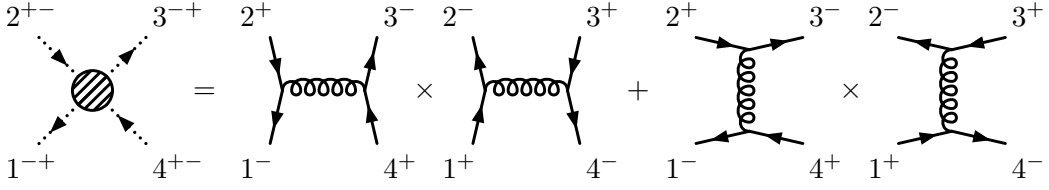}
\caption[a]{\small A double copy~\eqref{dilatonaxionexample} of diagrams
that obeys the color-kinematics duality in the fundamental representation.
It gives a four-scalar amplitude with gravitational interactions.}
\label{example4pt}
\end{figure}

There are many interesting generalizations of these amplitudes, including fermionic states,  but let us now focus on the amplitudes that will be relevant for obtaining pure Einstein gravity. We are interested in the ``impurities'' that come from dilaton and axion amplitudes. Since these particles are scalars, their amplitudes have no helicity weights, and the double copies should then be such that the external helicities are anticorrelated, as shown in \fig{example4pt}. The amplitude in that figure is given by
\be
   M_4^{\rm tree}(1^{-+}_\varphi\!,2^{+-}_\varphi\!,3^{-+}_\varphi\!,4^{+-}_\varphi)
      = -i \frac{n_s \overline{n}_s}{s}+i\frac{n_t \overline{n}_t}{t}
      = -i u^2 \Big(\frac{1}{s}+\frac{1}{t}\Big) = i\frac{u^3}{st} \,,
\label{dilatonaxionexample}
\ee
where, as before, $\overline{n}_i$ are the numerators in \eqn{qqQQ} with conjugated matter and antimatter.

Indeed, the states in the amplitude (\ref{dilatonaxionexample}) correspond to a complex scalar $\varphi^{-+}$ that can be understood as a linear combination of the dilaton and axion, $\varphi^{-+} \sim \phi+i a$, and similarly for the complex conjugate $\varphi^{+-} \sim \phi-i a$.
If we have two different complex scalars, $\varphi$ and $\varphi'$, then we can have the following amplitude by removing the $t$-channel graph:
\be
   M_4^{\rm tree}(1^{-+}_\varphi\!,2^{+-}_\varphi\!,3^{-+}_{\varphi'}\!,4^{+-}_{\varphi'})
      = -i \frac{n_s \overline{n}_s}{s}= -i\frac{u^2}{s} \,.
\label{twoscalarexample}
\ee

\begin{figure}[t]
      \centering
      \includegraphics[scale=1.00]{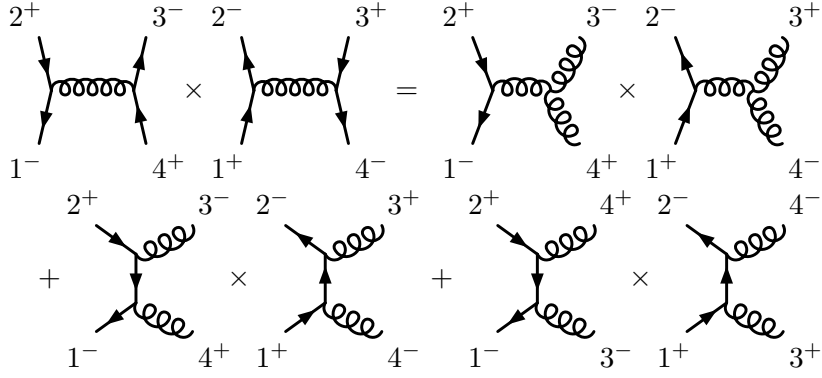}
\caption[a]{\small An equality between two different double copies: with two distinct fermion lines and with a single fermion line. Note that the propagators are implicitly included in this identity, and the $t$- and $u$-channel poles on the second line are spurious.}
\label{FDC4mixedFigure}
\end{figure}

As will be crucial for us in the following, these dilaton-axion amplitudes were constructed by taking double copies of quark states. This is not the conventional way, as most often we think of dilaton-axion amplitudes as arising in the double copy of YM amplitudes with only gluon states, corresponding to states with polarization tensors $ \varepsilon_{\mu\nu}^{+-} = \varepsilon_{\mu}^+ \varepsilon_{\nu}^- $,
$ \varepsilon_{\mu\nu}^{-+} = \varepsilon_{\mu}^- \varepsilon_{\nu}^+ $.
In particular, in the KLT formalism~\cite{Kawai:1985xq}, we would obtain the gravity amplitude in \eqn{dilatonaxionexample} as a product of color-ordered gluon amplitudes:
\be
   M_4^{\rm tree}(1^{-+}_\varphi\!,2^{+-}_\varphi\!,3^{-+}_\varphi\!,4^{+-}_\varphi)
      = -i s A_4^{\rm tree}(1^-_{g},2^+_{g},3^-_{g},4^+_{g})
             A_4^{\rm tree}(2^-_{g},1^+_{g},3^+_{g},4^-_{g})
      =  i \frac{u^3}{st} \,,
\label{KLT4example}
\ee
which gives the same result as our construction in \eqn{dilatonaxionexample}.
Similarly, one can build the amplitude in \eqn{twoscalarexample} through the double copy of gluons and a single fermion pair, either using the KLT prescription or the color-kinematics duality for adjoint and fundamental particles. The KLT formula gives
\be
   M_4^{\rm tree}(1^{-+}_\varphi\!,2^{+-}_\varphi\!,3^{-+}_{\varphi'}\!,4^{+-}_{\varphi'})
      = -i s A_4^{\rm tree}(1^-_{f},2^+_{f},3^-_{g},4^+_{g})
             A_4^{\rm tree}(2^-_{f},1^+_{f},3^+_{g},4^-_{g})
      = -i \frac{u^2}{s} \,,
\ee
which again agrees with \eqn{twoscalarexample}. Since the KLT approach is equivalent to the tree-level adjoint color-kinematics double copy, we have an interesting equality between the double copies of a single diagram (\ref{twoscalarexample}) and a triplet of diagrams, as is illustrated in \fig{FDC4mixedFigure}.

For amplitudes with a single fermion pair, one can take the fermions either in the adjoint or in the fundamental representation, and the color-kinematics duality and double-copy construction work in the same way for both cases. This, together with the identity in \fig{FDC4mixedFigure}, explains why we were able to obtain the four-point amplitudes from both the KLT approach and the fundamental color-kinematics duality. However, in general, at higher points, the KLT formula~\cite{Kawai:1985xq,Bern:1998sv} will not be able to reproduce the wide range of gravitational interactions that can be obtained from the double copy of diagrams satisfying the fundamental color-kinematics duality. This is because the latter approach should admit an unlimited number of distinguishable (chiral) matter particles (see \sect{numeratorDoubleCopy}), whereas the KLT formula should be limited to the double-copy spectrum of the adjoint states naturally occurring in SYM theories.

\subsubsection{Five-point tree amplitudes}
\label{tree5pt}

The ease of constructing four-point gravity amplitudes is no coincidence. Let us add another particle to the picture, a gluon, and observe how the double-copy structure generalizes. For easier bookkeeping, we first assume that the quarks have distinct flavors,
so the process is of the following type: $q\bar{q}q'\bar{q}'g$.
There are only five nonvanishing tree-level diagrams that can be constructed from Feynman rules. For completeness, we give all five contributions:
\begin{subequations} \begin{align}
      \parbox{91pt}{
      \begin{fmffile}{qqQQg1} \fmfframe(10,17)(0,17){
      \fmfsettings
      \begin{fmfgraph*}(60,40)
            \fmflabel{$\!\!1^-\!, i$}{q1}
            \fmflabel{$2^+\!, \bar \jmath\!\!\!\!\!$}{q2}
            \fmflabel{$3^-\!, k\!\!\!\!\!$}{q3}
            \fmflabel{$\!\!4^+\!, \bar l$}{q4}
            \fmflabel{$\!5, a$}{g5}
            \fmfleft{q2,q3}
            \fmfright{q1,g5,q4}
            \fmf{plain_arrow}{q2,v1,v5,q1}
            \fmf{plain_arrow}{q4,v3,q3}
            \fmf{curly,tension=0}{v3,v1}
            \fmf{curly,tension=0}{g5,v5}
      \end{fmfgraph*} }
      \end{fmffile}
      }
      & = \frac{i}{\sqrt{2}} \frac{1}{s_{15}s_{34}} \,
          T_{i \bar m}^a T_{m \bar \jmath}^b T_{k \bar l}^b \,
          \braket{1|\varepsilon_5|1\!+\!5|3} [24] = -i \frac{c_1 n_1}{D_1} \,,
\label{qqQQg1} \\
      \parbox{91pt}{
      \begin{fmffile}{qqQQg2} \fmfframe(10,17)(0,17){
      \fmfsettings
      \begin{fmfgraph*}(60,40)
            \fmflabel{$\!\!1^-\!, i$}{q1}
            \fmflabel{$2^+\!, \bar \jmath\!\!\!\!\!$}{q2}
            \fmflabel{$3^-\!, k\!\!\!\!\!$}{q3}
            \fmflabel{$\!\!4^+\!, \bar l$}{q4}
            \fmflabel{$5, a\!$}{g5}
            \fmfleft{q2,g5,q3}
            \fmfright{q1,q4}
            \fmf{plain_arrow}{q2,v5,v1,q1}
            \fmf{plain_arrow}{q4,v3,q3}
            \fmf{curly,tension=0}{v1,v3}
            \fmf{curly,tension=0}{v5,g5}
      \end{fmfgraph*} }
      \end{fmffile}
      }
      & = - \frac{i}{\sqrt{2}} \frac{1}{s_{25}s_{34}} \,
          T_{i \bar m}^b T_{m \bar \jmath}^a T_{k \bar l}^b \,
          \braket{13} [2|\varepsilon_5|2\!+\!5|4] = -i \frac{c_2 n_2}{D_2} \,,
\label{qqQQg2} \\
      \parbox{91pt}{
      \begin{fmffile}{qqQQg3} \fmfframe(10,17)(0,17){
      \fmfsettings
      \begin{fmfgraph*}(60,40)
            \fmflabel{$\!\!1^-\!, i$}{q1}
            \fmflabel{$2^+\!, \bar \jmath\!\!\!\!\!$}{q2}
            \fmflabel{$3^-\!, k\!\!\!\!\!$}{q3}
            \fmflabel{$\!\!4^+\!, \bar l$}{q4}
            \fmflabel{$5, a\!$}{g5}
            \fmfleft{q2,g5,q3}
            \fmfright{q1,q4}
            \fmf{plain_arrow}{q2,v1,q1}
            \fmf{plain_arrow}{q4,v3,v5,q3}
            \fmf{curly,tension=0}{v1,v3}
            \fmf{curly,tension=0}{g5,v5}
      \end{fmfgraph*} }
      \end{fmffile}
      }
      & = \frac{i}{\sqrt{2}} \frac{1}{s_{12}s_{35}} \,
          T_{i \bar \jmath}^b T_{k \bar m}^a T_{m \bar l}^b \,
          \braket{1|3\!+\!5|\varepsilon_5|3} [24] = -i \frac{c_3 n_3}{D_3} \,,
\label{qqQQg3} \\
      \parbox{91pt}{
      \begin{fmffile}{qqQQg4} \fmfframe(10,17)(0,17){
      \fmfsettings
      \begin{fmfgraph*}(60,40)
            \fmflabel{$\!\!1^-\!, i$}{q1}
            \fmflabel{$2^+\!, \bar \jmath\!\!\!\!\!$}{q2}
            \fmflabel{$3^-\!, k\!\!\!\!\!$}{q3}
            \fmflabel{$\!\!4^+\!, \bar l$}{q4}
            \fmflabel{$\!5, a$}{g5}
            \fmfleft{q2,q3}
            \fmfright{q1,g5,q4}
            \fmf{plain_arrow}{q2,v1,q1}
            \fmf{plain_arrow}{q4,v5,v3,q3}
            \fmf{curly,tension=0}{v3,v1}
            \fmf{curly,tension=0}{v5,g5}
      \end{fmfgraph*} }
      \end{fmffile}
      }
      & = - \frac{i}{\sqrt{2}} \frac{1}{s_{12}s_{45}} \,
          T_{i \bar \jmath}^b T_{k \bar m}^b T_{m \bar l}^a \,
          \braket{13} [2|4\!+\!5|\varepsilon_5|4] = -i \frac{c_4 n_4}{D_4} \,,
\label{qqQQg4} \\
      \parbox{91pt}{
      \begin{fmffile}{qqQQg5} \fmfframe(10,17)(0,17){
      \fmfsettings
      \begin{fmfgraph*}(60,40)
            \fmflabel{$\!\!1^-\!, i$}{q1}
            \fmflabel{$2^+\!, \bar \jmath\!\!\!\!\!$}{q2}
            \fmflabel{$3^-\!, k\!\!\!\!\!$}{q3}
            \fmflabel{$\!\!4^+\!, \bar l$}{q4}
            \fmflabel{$\!5, a$}{g5}
            \fmfleft{q2,q3}
            \fmfright{q1,g5,q4}
            \fmf{plain_arrow}{q2,v1,q1}
            \fmf{plain_arrow}{q4,v3,q3}
            \fmf{curly,tension=0.01}{v1,v5}
            \fmf{curly,tension=0.01}{v3,v5}
            \fmf{curly,tension=0}{g5,v5}
      \end{fmfgraph*} }
      \end{fmffile}
      } & = \!
      \begin{aligned} \\
          \frac{i}{\sqrt{2}} \frac{1}{s_{12}s_{34}} \,
          \tilde{f}^{abc} T_{i \bar \jmath}^b T_{k \bar l}^c \,
          \Big( \bra{1}\varepsilon_5|2] \bra{3}5|4]
              - \bra{1}5|2] \bra{3}\varepsilon_5|4] & \\
              -\,2 \braket{13} [24] ((k_1\!+\!k_2)\!\cdot\!\varepsilon_5) & \Big)
        = -i \frac{c_5 n_5}{D_5} \,.
      \end{aligned}\!\!\!\!\!\!\!\!\!\!\!\!\!\!\!\!\!\!\!\!
\label{qqQQg5}
\end{align} \label{qqQQg}%
\end{subequations}
The amplitude is the sum of these five diagrams.

In \eqn{qqQQg}, we defined five color factors $c_i$
that satisfy two Lie algebra identities:
\begin{subequations} \begin{align}
   c_1 - c_2 = - c_5 ~~~~~ \Leftrightarrow ~~~~~
   T_{i \bar m}^a T_{m \bar \jmath}^b T_{k \bar l}^b -
   T_{i \bar m}^b T_{m \bar \jmath}^a T_{k \bar l}^b & = -
   \tilde{f}^{abc} T_{i \bar \jmath}^b T_{k \bar l}^c \,, \\
   c_3 - c_4 = c_5 ~~~~~~\:\: \Leftrightarrow ~~~~~
   T_{i \bar \jmath}^b T_{k \bar m}^a T_{m \bar l}^b -
   T_{i \bar \jmath}^b T_{k \bar m}^b T_{m \bar l}^a & =
   \tilde{f}^{abc} T_{i \bar \jmath}^b T_{k \bar l}^c \,.
\end{align}%
\end{subequations}
For the amplitude to satisfy the color-kinematics duality, it is necessary that the kinematic numerators $n_i$ obey the corresponding identities $n_1 - n_2 = - n_5$ and $n_3 - n_4 = n_5$. As we are not yet probing the four-gluon vertex, it turns out that these identities automatically hold for the above Feynman diagrams, provided the gluon polarization vector is transverse.
This can be verified using some spinor algebra involving Schouten identities, for instance:
\be
   \braket{1|\varepsilon_5|1\!+\!5|3} [24]
   + \braket{13} [2|\varepsilon_5|2\!+\!5|4]
   = - \bra{1}\varepsilon_5|2] \bra{3}5|4]
     + \bra{1}5|2] \bra{3}\varepsilon_5|4] 
     +\,2 \braket{13} [24] ((k_1\!+\!k_2)\!\cdot \varepsilon_5) \,.
\ee

This five-point amplitude is an interesting example of the interplay between the color-kinematics duality with fundamental particles and the (generalized) gauge invariance. Note that the appearance of the kinematical Lie algebra identities is correlated with the fact that the five-point numerators are allowed to be gauge dependent. We can see this by parameterizing the polarization vectors with a reference momentum $q^\mu$. If we specialize to the case of the plus-helicity gluon, so that
$\varepsilon^\mu_{k+} = \bra{q}\gamma^\mu|k] / (\sqrt{2} \braket{qk})$,
we obtain
\begin{subequations} \begin{align}
   n_1 & = - \braket{13} [24] \frac{\bra{q}1|5]}{\braket{q5}} , \\
   n_2 & = \,\braket{13} \Big( [24] \frac{\bra{q}2|5]}{\braket{q5}}
                               + [25][54] \Big) , \\
   n_3 & = - \braket{13} [24] \frac{\bra{q}3|5]}{\braket{q5}} , \\
   n_4 & = \,\braket{13} \Big( [24] \frac{\bra{q}4|5]}{\braket{q5}}
                               - [25][54] \Big) , \\
   n_5 & = - n_1 + n_2 = n_3 - n_4 \,.
\end{align} \label{qqQQgnumerators}%
\end{subequations}
These numerators explicitly carry a nontrivial dependence on $q^\mu$.

Note that if all four quarks had the same flavor, $q\bar{q}q\bar{q} g$, as in \sect{tree4pt},
we would have to include five more diagrams with fermion lines
stretching between $4^+$ and $1^-$ and between $2^+$ and $3^-$.
Not surprisingly, they can be obtained from the diagrams in \eqn{qqQQg}
by relabeling $1\leftrightarrow3$.
However, by the two-term identity (\fig{TwoTermIdFigure}),
four of these numerators can be directly identified with those in \eqn{qqQQgnumerators}:
\beal
   n_6 & \equiv - n_1\big|_{1\leftrightarrow3} = n_3, \\ 
   n_7 & \equiv - n_2\big|_{1\leftrightarrow3} = n_2, \\
   n_8 & \equiv - n_3\big|_{1\leftrightarrow3} = n_1, \\
   n_9 & \equiv - n_4\big|_{1\leftrightarrow3} = n_4, \\
   n_{10} & \equiv - n_5\big|_{1\leftrightarrow3} ,
\label{TwoTermId5pt}
\eeal
where the negative signs come from swapping the fermions $1\leftrightarrow3$.
By combining the Lie algebra identities and the two-term identities,
one can then also express $n_{10}$ in terms of the previously given numerators,
$n_{10}= - n_3 + n_2 = n_1 - n_4$. 

Now, let us construct gravity amplitudes. As before, we can easily construct a four-photon one-graviton amplitude
\be
   M_5^{\rm tree}
      (1^{-}_\gamma,2^{+}_\gamma,3^{-}_\gamma,4^{+}_\gamma,5^{++}_h) =
      -i \sum_{i=1}^{10} \frac{n_i^2}{D_i} \,,
\label{5ptPhoton}
\ee
as well as a four-scalar one-graviton amplitude
\be
   M_5^{\rm tree}
      (1^{-+}_\varphi,2^{+-}_\varphi,3^{-+}_\varphi,4^{+-}_\varphi,5^{++}_h) =
      -i \sum_{i=1}^{10} \frac{n_i \overline{n}_i}{D_i} \,.
\label{FDC5}
\ee

Moreover, one can verify that,
in analogy with \eqns{TwoPhotonexample}{twoscalarexample}, taking just the five first diagrams~\eqref{qqQQg} results in the gauge-invariant amplitude that corresponds to gravitational interactions of two distinguishable scalars:
\be
   M_5^{\rm tree}
      (1^{-+}_\varphi,2^{+-}_\varphi,3^{-+}_{\varphi'},4^{+-}_{\varphi'},5^{++}_h) =
      -i \sum_{i=1}^{5} \frac{n_i \overline{n}_i}{D_i} \,.
\label{FDC5mixed}
\ee
The same procedure can be done for the photon amplitude~\eqref{5ptPhoton}.

At five points, the number of distinct matter particles is still quite small,
so we can reproduce the above amplitudes by applying
the KLT relation~\cite{Kawai:1985xq} in different ways:
\beal
   M_5^{\rm tree}(1,2,3,4,5)
      = i \big( s_{12} s_{34} A_5^{\rm tree}(1,2,3,4,5) & A_5^{\rm tree}(2,1,4,3,5) \\
         +\,s_{13} s_{24} A_5^{\rm tree}(1,3,2,4,5) & A_5^{\rm tree}(3,1,4,2,5) \big) \,.
\label{KLT5}
\eeal
Using this formula, one can easily check
that eqs.~\eqref{5ptPhoton}, \eqref{FDC5} and~\eqref{FDC5mixed} are indeed true.

Having considered these explicit tree-level examples of the color-kinematics duality for fundamental matter, we can now proceed to discussing formal aspects of how we treat supersymmetric YM theories with fundamental matter multiplets.

\subsection{On-shell spectrum and supermultiplets}
\label{supermultiplets}

The goal of this section is to package on-shell states of SYM theories
into their supermultiplets:
a vector multiplet ${\cal V}_{\cal N}$, as well as a chiral $ \Phi_{\cal N}$
and an antichiral $\overline{\Phi}_{\cal N}$ matter multiplet.
They respectively belong to the adjoint, fundamental
and antifundamental representations of the gauge group.

From now on, we use the same letters for on-shell states and their fields,
e.g., $A^\pm$ for the helicity states of gluons.
Moreover, we denote on-shell Weyl fermions by $\lambda^\pm$ or $\psi^\pm$,
depending on whether they belong to a vector or a matter multiplet, respectively. 
Similarly, the vector-multiplet scalars are denoted by $\varphi$
and the matter-multiplet ones by $\phi$.

Perhaps the best starting point for understanding the spectrum of massless gauge theories
is the on-shell supermultiplet of ${\cal N}=4$ SYM \cite{Kunszt:1985mg,Nair:1988bq}.
Using standard auxiliary Grassmann variables $\eta^A$,
this vector multiplet can be expressed as a super-wave function
\be
      V_{{\cal N}=4} = A^+ + \lambda_A^+ \eta^A+ \frac{1}{2} \varphi_{AB} \eta^A\eta^B
                     + \frac{1}{3!} \epsilon_{ABCD}\lambda_-^A\,\eta^B\eta^C\eta^D
                     + A_- \eta^1\eta^2\eta^3\eta^4 \,,
\ee
where the R-symmetry indices transform in the fundamental representation of SU(4),
and $\epsilon_{ABCD}$ is the Levi-Civita tensor.
The ${\cal N}=4$ vector multiplet is by necessity non-chiral
in contrast with the lower supersymmetry cases. Nevertheless, lower supersymmetric multiplets are direct truncations of this one. 

The on-shell vector multiplets for reduced supersymmetry are divided into
a chiral multiplet $V_{\cal N}$ and an antichiral one $\overline{V}_{\cal N}$.
For ${\cal N}=2,1,0$ they are explicitly:
\beal
      V_{{\cal N}=2} & = A^+ + \lambda^+_A \eta^A + \varphi_{12} \eta^1 \eta^2 \,,
      ~~~~~~
      \overline{V}_{{\cal N}=2} = \overline{\varphi}^{12} 
                                + \epsilon_{AB} \lambda_-^A \eta^B
                                + A_- \eta^1 \eta^2 \,, \\
      V_{{\cal N}=1} & = A^+ + \lambda^+ \eta^1 \,,
      ~~~~~~~~~~~~~~~~~~~~\,
      \overline{V}_{{\cal N}=1} = \lambda_- + A_- \eta^1 \,, \\
      V_{{\cal N}=0} & = A^+ \,,
      ~~~~~~~~~~~~~~~~~~~~~~~~~~~~~~\,
      \overline{V}_{{\cal N}=0} = A_- \,,
\label{chiralVectorMult}
\eeal
where SU(2) indices $A,B=1,2$ are inherited from SU(4) R-symmetry
and are raised and lowered using $\epsilon_{AB}$.
We find it convenient to assemble these chiral vector multiplets
into a single non-chiral multiplet
\be
      {\cal V}_{\cal N} = V_{\cal N} + \overline{V}_{\cal N} \, \theta \,,
\label{nonChiralMult}
\ee
where we introduced an auxiliary parameter $\theta$ defined as
\be
      \theta = \!\!\! \prod_{A={\cal N}+1}^4 \!\!\! \eta^A \,.
\label{theta}
\ee
It is nilpotent, $\theta^2=0$, and commuting or anticommuting
for an even or odd number of supersymmetries, respectively.

While the non-chiral multiplets \eqref{nonChiralMult} do not increase the supersymmetry
with respect to their chiral constituents,
they do make it possible for a more uniform treatment of the components
of the ${\cal N}=2,1,0$ scattering amplitudes
by assembling its full state dependence into a single generating function
$ A(k_i, \eta^A_i, \theta_i) $.
For example, the $m$-point MHV tree amplitude in ${\cal N}<4$ SYM theories
can be written as
\be
   A^{\rm MHV, tree}_m(k_i, \eta^A_i, \theta_i)
      = i \frac{ \delta^{(2{\cal N})}(Q)
                 \sum_{i<j}^m \theta_{i} \theta_{j} \spa{i}.{j}^{4-{\cal N}} }
               { \spa{1}.{2} \spa{3}.{4} \dots \spa{m}.{1} } \,,
\label{MHVampl}
\ee
where $\theta$'s mark the external legs that belong
to the $\overline{V}_{\cal N}$ multiplet
and $\eta$'s encode the on-shell supersymmetry inside both multiplets.
We define the delta function of the supermomentum
$ Q^{A}_\alpha = \sum_i |i\rangle_\alpha \eta_i^A $ as 
\be
   \delta^{(2{\cal N})}(Q)
      = (-1)^{\lceil {\cal N}/2 \rceil} \,  \prod_{A=1}^{\cal N} \sum_{i<j}^m \eta^A_i \spa{i}.{j}  \eta^A_j \,,
\ee
where a sign factor is included so that the gluon components in \eqn{MHVampl} have standard normalizations in the ${\cal N}=1,2$ cases.

All vector multiplets discussed so far, whether chiral or non-chiral,
belong to the adjoint representation of the gauge group.
For the matter content, on the other hand,
one can choose between the adjoint, fundamental or antifundamental representations.
The last two choices naturally define
the chiral matter multiplet $\Phi_{\cal N}$
and the antichiral one $\overline{\Phi}_{\cal N}$.
For ${\cal N}=2,1$ they are explicitly
\beal
   (\Phi_{{\cal N}=2})_i &= \psi^+_i+\phi_{Ai} \eta^A + \psi^{-}_i \eta^1 \eta^2 \,,
   ~~~~~~
   (\overline{\Phi}_{{\cal N}=2})_{\bar \imath}
      = \psi^+_{\bar \imath} + \epsilon_{AB}\,\overline{\phi}^A_{\bar \imath} \eta^B
                             + \psi^{-}_{\bar \imath} \eta^1 \eta^2 \,, \\
   (\Phi_{{\cal N}=1})_i &= \psi^+_i+\phi_i\,\eta^1 \,,
   ~~~~~~~~~~~~~~~~~~~~\,\,
   (\overline{\Phi}_{{\cal N}=1})_{\bar \imath}
      = \overline{\phi}_{\bar \imath} + \psi^-_{\bar \imath} \eta^1 \,.
\label{chiralMatterMult}
\eeal
Note that although the two ${\cal N}=2$ matter multiplets have similar states,
they are distinct and belong to different group representations,
as emphasized by the explicit color indices.

Non-supersymmetric matter can either be a chiral Weyl fermion or a complex scalar:
\beal
      (\Phi_{{\cal N}=0})_i \equiv     (\Phi_{{\cal N}=0}^{\rm fermion})_i &= \psi^+_i \,,~~~~~~~~~~~~
      (\overline{\Phi}_{{\cal N}=0})_{\bar \imath}\equiv       (\overline{\Phi}_{{\cal N}=0}^{\rm fermion})_{\bar \imath}
            = \psi^-_{\bar \imath} \,, \\
      (\Phi_{{\cal N}=0}^{\rm scalar})_i &= \phi_i \,,\hskip 3.85cm
      (\overline{\Phi}_{{\cal N}=0}^{\rm scalar})_{\bar \imath}
            = \overline{\phi}_{\bar \imath} \,.
\eeal
We take Weyl fermions to be the default ${\cal N}=0$ matter multiplets.
With this choice, the formalism for constructing pure supergravities will be uniform for any value of ${\cal N}$.

For completeness and for later use, we remark that
if the chiral and antichiral ${\cal N}=1$ multiplets~\eqref{chiralMatterMult}
are put in the adjoint representation of the gauge group
(or any real representation)
it is convenient to combine them into
a non-chiral minimal ${\cal N}=2$ multiplet.\footnote{Indeed,
this matter multiplet is often called ``${\cal N}=1$ chiral''
in the literature on scattering amplitudes.
To avoid confusion with the fundamental chiral multiplets,
here we prefer to call it ``${\cal N}=2$ adjoint.'' Furthermore, for simplicity, we choose to not use the word ``hypermultiplet''.}
After relabeling the scalar fields, we have
\be
      \Phi_{{\cal N}=1} + \overline{\Phi}_{{\cal N}=1} \, \eta^2
      ~~~~ \rightarrow ~~~~
      \Phi_{{\cal N}=2}^{\rm adj}=\psi^++\phi_{A}\eta^A+\psi^-\eta^1\eta^2 \,.
\label{adjointNeq2}
\ee
Likewise,
the non-supersymmetric chiral fermion matter can be promoted to a Majorana fermion,
and the fundamental complex scalar becomes equivalent to two real scalars,
all in  the adjoint representation.

Finally, in \tab{Multiplets} we collect the particle and helicity content of all discussed YM supermultiplets. The explicit interactions of the fundamental multiplets are summarized in \app{AppendixA}.

\begin{table*}
\centering
\begin{tabular}{|c|c|c|c|c|c||c|c|c|c||c|c|c|c|}
\hline 
Field & $+1$ & $\!+\frac{1}{2}$ & $\,\,0\,\,$ & $\!-\frac{1}{2}$ & $-1$ &
Field & $\!+\frac{1}{2}$ & $\,\,0\,\,$ & $\!-\frac{1}{2}$ &
Field & $\!+\frac{1}{2}$ & $\,\,0\,\,$ & $\!-\frac{1}{2}$ \\
\hline 
$V_{{\cal N}=4}$ &1&4&6&4&1 & $\Phi_{{\cal N}=2}$&1&2&1 & $\overline{\Phi}_{{\cal N}=2}$ &1&2&1 \\
\hline 
${\cal V}_{{\cal N}=2}$ &1&2&2&2&1& $\Phi_{{\cal N}=1}$&1&1&0 & $\overline{\Phi}_{{\cal N}=1}$ &0&1&1 \\
\hline 
${\cal V}_{{\cal N}=1}$ &1&1&0&1&1& $\Phi_{{\cal N}=0}$ &1&0&0& $\overline{\Phi}_{{\cal N}=0}$&0&0&1 \\
\hline
${\cal V}_{{\cal N}=0}$ &1&0&0&0&1& $\Phi^{\rm scalar}_{{\cal N}=0}$ &0&1&0&  $\Phi^{\rm adj}_{{\cal N}=2}$ &1&2&1 \\
\hline
\end{tabular}
\caption[a]{\small Summary of the particle and helicity content of the various on-shell YM supermultiplets considered in this paper.}
\label{Multiplets} 	
\end{table*}

\section{Construction of pure gravity theories}
\label{FundDoubleCopy}

In this section, we address the problem of constructing pure (super-)gravity theories with ${\cal N}<4$ supersymmetry. Such theories are ``non-factorizable'', meaning that their loop amplitudes cannot be constructed by squaring, or double copying, numerators of pure (super-)Yang-Mills theories. However, remarkably, pure gravity amplitudes can be obtained from non-pure YM theory, as we show next.

\subsection{Double copies of physical states}
\label{tensorrules}

We define the following tensor products of the adjoint and fundamental on-shell states in SYM theories with ${\cal N}$- and ${\cal M}$-extended supersymmetry:
\begin{subequations} \begin{align}
   \text{factorizable graviton multiplet}: \qquad
      {\cal H}_{{\cal N}+{\cal M}} & \equiv
      {\cal V}_{\cal N} \otimes {{\cal V}'\!}_{\cal M}\,,
\label{tensoringvector} \\
   \text{gravity matter}: \qquad
      {X}_{{\cal N}+{\cal M}} & \equiv
      \Phi_{\cal N} \otimes \overline{\Phi}'_{\cal M}\,,
\label{tensoringmatter}  \\
   \text{gravity antimatter}: \qquad
      \overline{X}_{{\cal N}+{\cal M}} & \equiv
      \overline{\Phi}_{\cal N} \otimes \Phi'_{\cal M}\,,
\label{tensoringantimatter} 
\end{align} \label{supertensoring}%
\end{subequations}
where we take $0 \le {\cal N} \le 2$ and $0 \le {\cal M} \le 2$,
so that $X_{{\cal N}+{\cal M}}$ is either a vector multiplet $V_{{\cal N}+{\cal M}}$,
or a lower-spin matter multiplet $\Phi_{{\cal N}+{\cal M}}$.
These are the states that naturally appear in the double copies of amplitudes
that obey the color-kinematics duality for adjoint and fundamental particles.

\begin{table*}
\centering
\begin{tabular}{|c|c|c|}
\hline 
SUGRA & tensoring vector states &
$ \text{ghosts} = \text{matter} \otimes \overline{\text{matter}}$ \phantom{\big[} \\
\hline
\;\!\!\!${\cal N}\!=0\;\!\!+\;\!\!0$\!\!\!\; &
$A^\mu \;\!\!\otimes\:\!\! A^\nu \;\!\! = \;\!\! h^{\mu \nu} \oplus \phi \oplus a$ &
$ (\psi^+\!\otimes\:\!\! \psi^-) \oplus (\psi^-\!\otimes\:\!\! \psi^+)
            = \phi \oplus a $ \phantom{\Big[}\!\! \\
\hline
\;\!\!\!${\cal N}\!=1\;\!\!+\;\!\!0$\!\!\!\; &
${\cal V}_{{\cal N}=1} \;\!\!\otimes\:\!\! A^\mu \;\!\!
      = \;\!\! H_{{\cal N}=1} \oplus \Phi_{{\cal N}=2}$ &
$ (\Phi_{{\cal N}=1} \;\!\!\otimes\:\!\! \psi^{-})  \oplus
            (\overline{\Phi}_{{\cal N}=1} \;\!\!\otimes\:\!\! \psi^{+})
            = \Phi_{{\cal N}=2} $ \phantom{\Big[}\!\!\! \\
\hline
\;\!\!\!${\cal N}\!=2\;\!\!+\;\!\!0$\!\!\!\; &
${\cal V}_{{\cal N}=2} \;\!\!\otimes\:\!\! A^\mu \;\!\!
      = \;\!\! H_{{\cal N}=2} \oplus {\cal V}_{{\cal N}=2}$ &
$ (\Phi_{{\cal N}=2} \;\!\!\otimes\:\!\! \psi^{-}) \oplus
      (\overline{\Phi}_{{\cal N}=2} \;\!\!\otimes\:\!\! \psi^{+})
      = {\cal V}_{{\cal N}=2} $ \phantom{\Big[}\!\! \\
\hline
\;\!\!\!${\cal N}\!=1\;\!\!+\;\!\!1$\!\!\!\; &
\!${\cal V}_{{\cal N}=1} \;\!\!\otimes\! {\cal V}_{{\cal N}=1} \;\!\!
      = \;\!\! H_{{\cal N}=2} \oplus 2\Phi_{{\cal N}=2}$\!\! &
\!$ (\Phi_{{\cal N}=1} \;\!\!\otimes\:\!\! \overline{\Phi}_{{\cal N}=1}) \oplus
      (\overline{\Phi}_{{\cal N}=1} \;\!\!\otimes\:\!\! \Phi_{{\cal N}=1})
      = 2 \Phi_{{\cal N}=2} $ \phantom{\Big[}\!\!\!\!\!\!\!\!\; \\
\hline
\;\!\!\!${\cal N}\!=2\;\!\!+\;\!\!1$\!\!\!\; &
\!${\cal V}_{{\cal N}=2} \;\!\!\otimes\! {\cal V}_{{\cal N}=1} \;\!\!
      = \;\!\! H_{{\cal N}=3} \oplus {\cal V}_{{\cal N}=4}$~\,\!\! &
\!$ (\Phi_{{\cal N}=2} \;\!\!\otimes\:\!\! \overline{\Phi}_{{\cal N}=1}) \oplus
      (\overline{\Phi}_{{\cal N}=2} \;\!\!\otimes\:\!\! \Phi^+_{{\cal N}=1})
      = {\cal V}_{{\cal N}=4} $ \phantom{\Big[}\!\!\! \\
\hline
\;\!\!\!${\cal N}\!=2\;\!\!+\;\!\!2$\!\!\!\; &
\!${\cal V}_{{\cal N}=2} \;\!\!\otimes\! {\cal V}_{{\cal N}=2} \;\!\!
      = \;\!\! H_{{\cal N}=4} \oplus 2{\cal V}_{{\cal N}=4}$\!\! &
\!$ (\Phi_{{\cal N}=2} \;\!\!\otimes\:\!\! \overline{\Phi}_{{\cal N}=2}) \oplus
      (\overline{\Phi}_{{\cal N}=2} \;\!\!\otimes\:\!\! \Phi_{{\cal N}=2})
      = 2 {\cal V}_{{\cal N}=4} $ \phantom{\Big[}\!\!\!\!\!\! \\
\hline
\end{tabular}
\caption[a]{\small Pure gravities are constructed from states that are double copies of pure SYM vectors, and similarly ghosts from matter-antimatter double copies. For compactness, pairs of chiral vectors, or pairs of chiral matter multiplets, are combined into non-chiral real multiplets. The ${\cal N}=2+0$ and ${\cal N}=1+1$ cases should give alternative constructions of the same pure ${\cal N}=2$ supergravity.}
\label{DCconstructions}
\end{table*}

The factorizable graviton multiplet~\eqref{tensoringvector},
naturally obtained in the adjoint double copy~\eqref{BCJformGravityAdj},
is characteristic of some supergravity theory,
but it is not the spectrum of a pure supergravity theory.
Indeed, ${\cal H}$ is reducible into the (non-chiral) pure graviton multiplet $H$
and a (chiral) complex matter multiplet $X$:
\be
   {\cal H}_{{\cal N}+{\cal M}} \equiv {\cal V}_{\cal N} \otimes {{\cal V}'\!}_{\cal M}
      = H_{{\cal N}+{\cal M}} \oplus {X}_{{\cal N}+{\cal M}}
                              \oplus {\overline X}_{{\cal N}+{\cal M}} \,.
\label{tensordecomposition}
\ee
For example, in the bosonic case the double copy reduces to the graviton, dilaton and axion:
$A^\mu \otimes A^\nu = h^{\mu\nu} \oplus \varphi^{+-} \oplus \varphi^{-+} $
with $\varphi \sim \phi +i a$.
This and other cases are explicitly listed in \tab{DCconstructions}.
While this reduction is easy to carry out for the on-shell asymptotic states,
the same is not true for the off-shell internal states
that are obtained from the double-copy construction \eqref{BCJformGravityAdj}.
The reason is that the double copies of gluons $A^\mu \otimes A^\nu$ are difficult
to decompose since amplitudes have no free Lorentz indices,
or, alternatively, reducing the product using little-group indices
would break Lorentz invariance of the off-shell states.

Instead, we propose to use the fact that multiplets $X$ and $\overline{X}$ appear not only in the product~\eqref{tensordecomposition} of the adjoint vectors
but also as the fundamental matter double copy
in \eqns{tensoringmatter}{tensoringantimatter}.
This will let us take the factorizable graviton multiplet~${\cal H}$
and mod out by~$X$ and~$\overline{X}$.
To do that, we will promote the matter double copies $X$ and $\overline{X}$ to be ghosts, i.e. opposite-statistics states.\footnote{This is superficially similar to the Faddeev-Popov method~\cite{Faddeev:1967fc} that removes the unphysical YM states from propagating in loops, although the details are quite different.}

First of all, let us do some simple counting of the on-shell degrees of freedoms
to show that it is indeed the same $X$, $\overline{X}$
that appear in \eqns{supertensoring}{tensordecomposition}.
For this, we add the bosonic and fermionic counts of states in the multiplets.
To start with, note that all minimal (chiral) supermultiplets with ${\cal N}$-extended supersymmetry have exactly $2^{\cal N}$ states
(e.g. see \eqns{chiralVectorMult}{chiralMatterMult}) in four dimensions.
Therefore, non-chiral pure vector and graviton multiplets have twice as many states,
$2^{{\cal N}+1}$, except for the maximally-supersymmetric multiplets that we do not consider here.
For example, pure YM theory and Einstein gravity both have ${\cal N}=0$,
so they both contain $2^{0+1}=2$ physical states:
the two on-shell gluons or gravitons.
For this bosonic case, \eqn{tensordecomposition} becomes
\be
   4 = 2 \otimes 2 = 2 \oplus 1 \oplus 1 \,,
\ee
where the right-hand side represents the two gravitons, dilaton and axion,
or rather the mixed dilaton-axion states $\varphi^{+-}$ and $\varphi^{-+}$
introduced in \sect{tree4pt}. 
In that picture, $X$ can be thought of as $\varphi^{+-} = A^+ \otimes A^-$.
For a general $0\le{\cal N},{\cal M} \le 2$ supersymmetric theory,
\eqn{tensordecomposition} becomes
\be
   2^{{\cal N}+{\cal M}+2} = 2^{{\cal N }+1} \otimes 2^{{\cal M}+1}
                           = 2^{{\cal N}+{\cal M}+1} \oplus 2^{{\cal N}+{\cal M}}
                                                     \oplus 2^{{\cal N}+{\cal M}} \,.
\ee
Indeed, the right-hand side represents the pure graviton multiplet
in ${\cal N}+{\cal M}$ supergravity
plus two minimal (chiral) matter multiplets in the same theory.
Clearly, the matter multiplet $X$ defined in \eqn{tensoringmatter}
has the same state counting,
\be
   2^{{\cal N }+{\cal M }} =2^{{\cal N }} \otimes 2^{{\cal M }} \,,
\ee
and similarly for $\overline{X}$.
In the bosonic case, we have  $1=1\otimes 1$,
which now implies that $X=\varphi^{+-}=\psi^+ \otimes \psi^-$,
where $\psi^+$, $\psi^-$ are on-shell Weyl fermions.

In this way, we have shown that simple counting is consistent with our recipe for obtaining
the pure graviton multiplet $H$ from ${\cal V} \otimes {{\cal V}'}$
by modding out by $\Phi \otimes \overline{\Phi}'$ and $\overline{\Phi} \otimes \Phi'$.

\subsection{Numerator double copies}
\label{numeratorDoubleCopy}

In \sect{colorkinematics}, we considered (super-)Yang-Mills theories where adjoint and fundamental color representations are dual to vector and matter multiplets, respectively, in the sense that the kinematic structure of the amplitudes is governed by the adjoint and fundamental color-kinematics duality. In \sect{tensorrules}, we explained how the on-shell SYM states  can be tensored to obtain the on-shell spectrum of pure supergravities, including Einstein gravity. To obtain gravity amplitudes, we now have to do the same double-copy construction for the interacting theories with internal off-shell states.

Pure (super-)gravity amplitudes are obtained by recycling the cubic diagrams~\eqref{BCJformYM} of the YM theory and constructing the proper double copies of the kinematic numerators, similarly to \eqn{BCJformGravityAdj}. The numerator double copies should be constructed so that internal adjoint lines in gauge theory become an off-shell equivalent of ${\cal V} \otimes {\cal V}'$ lines in gravity. And similarly, fundamental and antifundamental gauge-theory lines become gravitational lines that are off-shell continuations of $\Phi \otimes \overline{\Phi}'$ and $\overline{\Phi} \otimes \Phi'$, respectively. That is, the numerator copies should {\it not} be taken between diagrams that produce cross terms between the vector and matter internal lines. Such double copies would in general not be consistent
since the kinematic algebra is different for the two types of states. 

In fact, the above double-copy structure is already present in YM theory. By definition, since \eqn{BCJformYM} describes gauge-theory amplitudes, there are no products of numerators and color factors, $n_i c_i$, involving cross-terms between internal vector kinematic lines and fundamental color lines. Thus the transition from YM to gravity amplitudes should be as straightforward as replacing the color factors with kinematic numerators. For example, replacing $c_i \rightarrow n_i'$ in \eqn{BCJformYM} would give valid gravity amplitudes, albeit with undesired matter content
given by $\Phi \otimes \Phi'$ and $\overline{\Phi} \otimes \overline{\Phi}'$.
To get matter states such as $\Phi \otimes \overline{\Phi}'$, we can instead let $c_i \rightarrow \overline{n}'_i$, where the bar operation swaps matter and antimatter. However, that would give gravity amplitudes with extra physical matter. Since we are trying to remove the matter already present in the decomposition of ${\cal V} \otimes {\cal V}'$, what we want is matter that behaves as ghosts. This means that we should insert a minus sign for each closed loop of matter particles. Hence we do the following replacement in \eqn{BCJformYM},
\be
   c_i \rightarrow (-1)^{|i|}\overline{n}'_i
\ee
where $|i|$ counts the number of closed matter loops in the $i$'th trivalent graph.
More generally, we can do the replacement $c_i \rightarrow (N_X)^{|i|} \overline{n}'_i$, where $N_X+1$ is the number of complex matter multiplets in the desired gravity theory.

Thus we propose the following formula for amplitudes in pure (super-)gravity:
\be
   {\cal M}^{\text{$L$-loop}}_m
      = i^{L-1} \Big(\frac{\kappa}{2}\Big)^{m+2L-2}
        \sum_{i} \int\!\!\frac{d^{LD}\ell}{(2\pi)^{DL}}
        \frac{(-1)^{|i|}}{S_i} \frac{n_i \overline{n}_i'}{D_i} \,,
\label{BCJformGravity}
\ee
where the sum runs over all $m$-point $L$-loop graphs with trivalent vertices
of two kinds: (fact. grav., fact. grav., fact. grav.) and (fact. grav., matter, antimatter),
corresponding to particle lines of two types: factorizable graviton multiplets and matter ghosts. 
The graphs are in one-to-one correspondence with the ones
in the gauge-theory amplitude~\eqref{BCJformYM},
and the denominators $D_i$ and the symmetry factors $S_i$ are same.
The objects $n_i$ and $n_i'$ are the numerators of two gauge-theory amplitudes
where at least one of the two copies obeys the color-kinematics duality.
The bar on top of one of the numerators denotes
the operation of conjugating the matter particles
and reversing the fundamental-representation flows.
Finally, the contribution of each graph enters the sum with a sign
determined by the number of closed matter-ghost loops $|i|$.

When there are no fundamental lines in the diagram,
both the matter conjugation the sign-modifying factor act trivially,
so eliminating all graphs with matter ghosts reduces \eqn{BCJformGravity}
to the adjoint double copy~\eqref{BCJformGravityAdj}.

\subsection{Gauge invariance of the gravity construction}
\label{GaugeInvarianceSect}

Now let us show that the double-copy construction~\eqref{BCJformGravity}
for pure gravity amplitudes obeys generalized gauge invariance.
This means that the freedom to shift numerators in gauge theory
should correspond to a similar freedom in shifting one copy of the gravity numerators.
If this was not true, the result of the double copy would depend on the gauge
chosen for the YM amplitude computation, and thus necessarily be wrong.
Moreover, while it is not obvious, experience tells us that the notion of generalized gauge invariance is perhaps as useful and constraining for amplitudes as the more familiar notion of gauge invariance is for Lagrangians. Therefore, showing that the gravity construction is generalized-gauge-invariant is a strong check of consistency.

Recall that the generalized gauge invariance in YM theory allows us to freely shift $n_i \rightarrow n_i + \Delta_i$ provided that $\Delta_i$ satisfy
\be
   \sum_{i} \int\!\!\frac{d^{LD}\ell}{(2\pi)^{DL}}
                    \frac{1}{S_i} \frac{ \Delta_i c_i }{D_i} =0\,.
\ee
The origin of this freedom comes from the overcompleteness of the color factor basis.
Indeed, the Jacobi identities and commutation relations satisfied by the color factors
can indirectly enter the amplitude multiplied by arbitrary kinematic functions,
which we may refer to as pure gauge terms.
For example, consider a shift of the numerators $n_i$, $n_j$ and $n_k$
of three diagrams
with identical graph structure except for a single edge:
\begin{subequations} \begin{align}
   {\cal A}^{\text{$L$-loop}}_m & \rightarrow {\cal A}^{\text{$L$-loop}}_m
      + 
        \int K_{\rm pure\,gauge}
            \big( \tf^{dac} \tf^{cbe} - \tf^{dbc} \tf^{cae} - \tf^{abc} \tf^{dce} \big)
            c^{abde}_{\rm rest} \,, \\
   {\cal A}^{\text{$L$-loop}}_m & \rightarrow {\cal A}^{\text{$L$-loop}}_m
      + 
        \int K_{\rm pure\,gauge}
            \big(T^{a}_{i \bar \jmath} \, T^{b}_{j \bar k}
               - T^{b}_{i \bar \jmath} \, T^{a}_{j \bar k}
               - \tf^{abc} \, T^{c}_{i \bar k} \big)
            c^{ab}_{k \bar \imath, {\rm rest}} \,,
\label{Ashift}
\end{align}%
\end{subequations}
where the standard loop integration measure is suppressed for brevity.
Taking into account that the symmetry factors and denominators
are included into $K_{\rm pure\,gauge}$,
the three numerators will be shifted by
\be
   \Delta_i = S_i D_i K_{\rm pure\, gauge} \,, ~~~
   \Delta_j = S_j D_j K_{\rm pure\, gauge} \,, ~~~
   \Delta_k = S_k D_k K_{\rm pure\, gauge} \,.
\ee
Of course, $K_{\rm pure\,gauge}$ is multiplied here by something that vanishes
due to the color algebra~\eqref{chiraljacobi},
so the amplitude cannot depend on it.

Translating this example to gravity
through the replacement $c_i \rightarrow (-1)^{|i|} \overline{n}'_i$, we obtain\footnote{
This argument trivially extends to the more general replacement prescription
$c_i \rightarrow (N_X)^{|i|} \overline{n}'_i$.}
\be
   M^{\text{$L$-loop}}_m \rightarrow M^{\text{$L$-loop}}_m
      + (-1)^{|i|} \int K_{\rm pure\,gauge}
        \big( {\overline{n}}'_i - {\overline{n}}'_j - {\overline{n}}'_k \big)  \,.
\label{Mshift}
\ee
The crucial point here is that the commutation and Jacobi relations in \fig{ChiralJacobiFigure}
leave the number of closed fundamental matter loops invariant,
so in \eqn{Mshift} we safely set $|i|=|j|=|k|$.
For the gravitational amplitude~\eqref{Mshift}
to be free of the pure gauge redundancy,
the numerators $\overline{n}'_i$ must satisfy the color-dual algebra: ${\overline{n}}_i' - {\overline{n}}_j' = {\overline{n}}_k' $.

In summary, shifting the gauge-theory numerators $n_i \rightarrow n_i + \Delta_i$
gives new gravity numerators $n_i \overline{n}'_i \rightarrow (n_i + \Delta_i) \overline{n}'_i$.
If $\overline{n}_i$ satisfy the same algebraic relations as the color factors~$c_i$,
then we are guaranteed to leave the gravity amplitude invariant,
since it inherits the YM identity~\eqref{YMshift} in the form
\be
\sum_{i} \int\!\!\frac{d^{LD}\ell}{(2\pi)^{DL}}
                   \frac{1}{S_i} \frac{ \Delta_i \overline{n}'_i }{D_i} =0\,.
\label{GRshift}
\ee
As in the purely-adjoint case~\cite{Bern:2010yg},
this argument can be easily extended to show that,
to obtain a consistent double-copy amplitude,
it is sufficient for only
one of the two numerator sets ($n_i$ or $\overline{n}'_i$)
to satisfy the color-kinematics duality, 

From this line of arguments, one can also see that
the additional two-term algebraic identity~\eqref{4ptTwoTerm},
depicted in \fig{TwoTermIdFigure}, is optional.
It is not needed for obtaining generalized-gauge-invariant amplitudes
since the corresponding color identity is not generically present for fundamental representations of, say, SU($N$).
In fact, in those spacetime dimensions where the two-term identity~\eqref{4ptTwoTerm}
can be imposed, the double-copy gravity amplitudes stay the same
with or without imposing it.
Indeed, one can think of the identity~\eqref{4ptTwoTerm}
as a special gauge choice in YM theory,
hence the gravity amplitudes cannot depend on this choice.
We have also confirmed this property by constructing explicit tree-level amplitudes
through seven points.

\subsection{Checks of two-loop cuts}
\label{cutchecks}

\begin{figure}[t]
      \centering
      \includegraphics[scale=1.00]{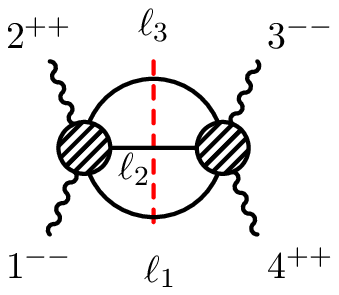}
      \hspace{30pt}
      \includegraphics[scale=1.00,trim=0 10pt 0 0]{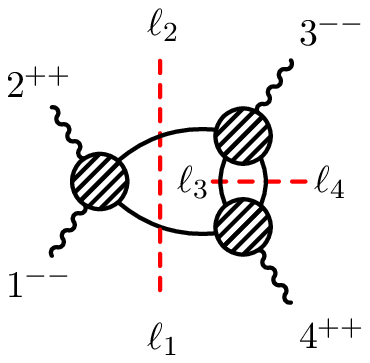}
\caption[a]{\small Two-loop cuts of the four-graviton amplitude
that can have ghost matter circulating in only a single loop at the time.}
\label{twoloopcut4ptAB}
\end{figure}

\begin{figure}[t]
      \centering
      \includegraphics[scale=1.00]{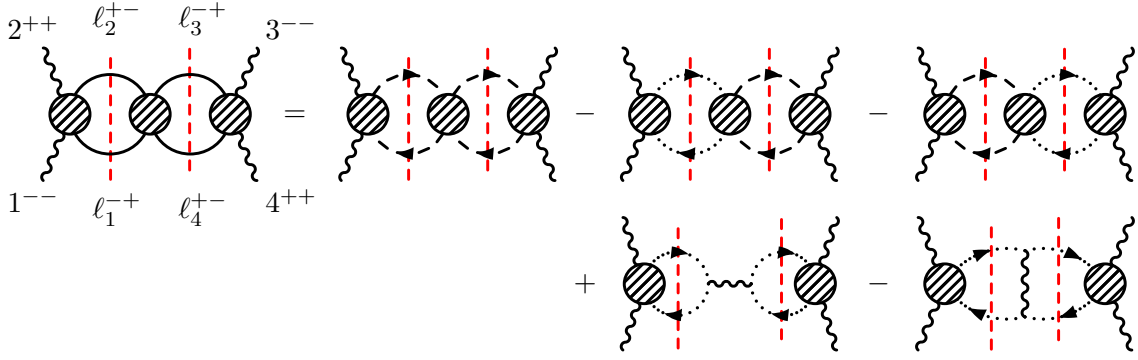}
      \vspace{-5pt}
\caption[a]{\small A two-loop cut of the four-graviton amplitude.
A particular internal helicity configuration is expanded using
the diagrammatic rules given in \fig{DoubleCopyLines}.
In the text, we explain that these dilaton-axion contributions
cancel among themselves, leaving the pure gravity cut.}
\label{twoloopcut4pt}
\end{figure}

In this section, we consider an explicit multiloop example that provides a nontrivial check that the double-copy prescription~\eqref{BCJformGravity} correctly eliminates the dilaton and axion contributions from the unitarity cuts in Einstein gravity. We concentrate on two-loop cuts in four dimensions; the one-loop case will be treated in full rigor in subsequent sections with complete four-point calculations in $D=4-2\epsilon$ dimensions. For the construction of the two-loop four-point amplitude it is sufficient to consider three types of unitarity cuts: the two shown in \fig{twoloopcut4ptAB} and the first diagram shown in \fig{twoloopcut4pt}. For the purpose of checking the cancellation of dilaton and axion contributions the latter cut is the most nontrivial. The two cuts in \fig{twoloopcut4ptAB} can only have matter circulating in a single sub-loop at the time, so they can at most teach us about the dilaton-axion cancelations that take place in one-loop amplitudes, the topic of the following sections.

\begin{figure}[t]
      \centering
      \includegraphics[scale=1.00]{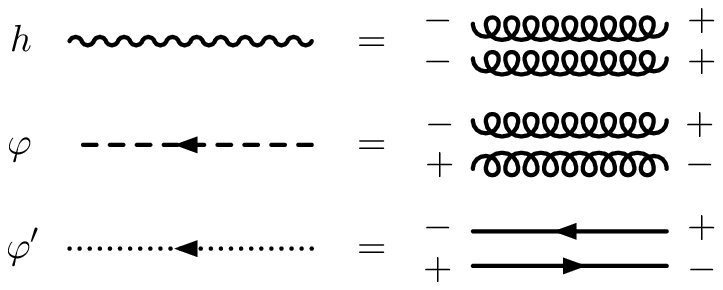}
\caption[a]{\small Diagrammatic rules for the gravity cut lines in \fig{twoloopcut4pt}:
graviton lines correspond to double copies of gluon lines,
whereas mixed dilaton-axion lines can be realized either in the same way
or as fundamental double copies of fermions.}
\label{DoubleCopyLines}
\end{figure}

Now, consider the double $s$-channel cut of the four-graviton two-loop amplitude. To be explicit, consider the external configuration $M(1^{--}_h\!, 2^{++}_h\!, 3^{--}_h\!, 4^{++}_h) $, and focus only on the internal-matter contributions as shown in \fig{twoloopcut4pt}. If the theory is pure, these matter contributions should add up to zero. As shown, the cut has an expansion in terms of different particle and diagram contributions coming from our prescription. In order to not clutter the diagrams with explicit double-line notation that reflect the tensor structure of the gravity states, we use the particle-line notation given in \fig{DoubleCopyLines}. As explained in \sect{tree4pt}, the complexified dilaton-axion contributions arise as double copies of gluons with anticorrelated helicities, gravitons from double copies of correlated gluons, and matter ghosts from anticorrelated fermions (quarks).

The diagrammatic expansion in \fig{twoloopcut4pt} is straightforward, except for the special treatment of the opposite-statistics nature of the ghosts. In particular, in the last two diagrams the intermediate graviton channel needs to be resolved before the sign of the diagram can be determined due to the different number of ghost loops.
 
To calculate the result of the cut, we can take advantage of a convenient simplification. 
If we focus for a moment on the leftmost and rightmost four-point tree amplitudes (outer blobs) appearing in the factorization of the cut, we see that they are in fact identical for all the five contributions in \fig{twoloopcut4pt}. They are explicitly
\begin{subequations} \begin{align}
\begin{aligned}
   M_4^{\rm tree}
      (1^{--}_h\!,2^{++}_h\!,(-\ell_2)^{-+}_{\varphi}\!,(-\ell_1)^{+-}_{\varphi})
&= M_4^{\rm tree}
      (1^{--}_h\!,2^{++}_h\!,(-\ell_2)^{-+}_{\varphi'}\!,(-\ell_1)^{+-}_{\varphi'}) \\
&= \frac{i \bra{1}\ell_1|2]^4}{s (k_1\!-\!\ell_1)^2 (k_1\!-\!\ell_2)^2} \,,
\end{aligned}
\label{exttree1} \\
\begin{aligned}
   M_4^{\rm tree}
      (3^{--}_h\!,4^{++}_h\!,(-\ell_4)^{+-}_{\varphi}\!,(-\ell_3)^{-+}_{\varphi})
&= M_4^{\rm tree}
      (3^{--}_h\!,4^{++}_h\!,(-\ell_4)^{+-}_{\varphi'}\!,(-\ell_3)^{-+}_{\varphi'}) \\
&= \frac{i \bra{3}\ell_3|4]^4}{s (k_3\!-\!\ell_3)^2 (k_3\!-\!\ell_4)^2} \,,
\end{aligned}
\label{exttree2}
\end{align} \label{exttrees} \end{subequations} 
\!\!where $\varphi$ denote the complexified dilaton-axion states arising from gluon double copies, and $\varphi'$ are the fundamental fermion double copies.
These amplitudes contribute to the cut in \fig{twoloopcut4pt} as an overall factor, and thus we may ignore them and only check the cancellation among the five terms associated with the gravity tree amplitudes in the middle blobs.
Conveniently, we already have all the ingredients at hand, from the results of \sect{tree4pt}.
The first term is  $M_4^{\rm tree} (\ell^{-+}_{1\,\varphi}\!,\ell^{+-}_{2\,\varphi}\!, \ell^{-+}_{3\,\varphi}\!,\ell^{+-}_{4\,\varphi})$,
and the second and third terms are identical:
\be
 - M_4^{\rm tree} (\ell^{-+}_{1\,\varphi'},\ell^{+-}_{2\,\varphi'},
                   \ell^{-+}_{3\,\varphi}\!,\ell^{+-}_{4\,\varphi}) = 
 - M_4^{\rm tree} (\ell^{-+}_{1\,\varphi}\!,\ell^{+-}_{2\,\varphi}\!,
                   \ell^{-+}_{3\,\varphi'},\ell^{+-}_{4\,\varphi'}) \,.
\ee
For the fourth and fifth terms we recycle the numerators $n_s$ and $n_t$ from \eqn{qqQQ}, with $k_i\rightarrow l_i$. Therefore, we get the following sum for the internal-matter cut:
\beal
     M_4^{\rm tree} (\ell^{-+}_{1\,\varphi}\!,\ell^{+-}_{2\,\varphi}\!,
                     \ell^{-+}_{3\,\varphi}\!,\ell^{+-}_{4\,\varphi})
 - 2 M_4^{\rm tree} (\ell^{-+}_{1\,\varphi}\!,\ell^{+-}_{2\,\varphi}\!,
                     \ell^{-+}_{3\, \varphi'},\ell^{+-}_{4\, \varphi'})
 - i \frac{n_s \overline{n}_s}{(\ell_1\!+\!\ell_2)^2}
 + i \frac{n_t \overline{n}_t}{(\ell_2\!+\!\ell_3)^2} & \\
 =-i \frac{n_s \overline{n}_s}{(\ell_1\!+\!\ell_2)^2}
 - i \frac{n_t \overline{n}_t}{(\ell_2\!+\!\ell_3)^2}
 +2i \frac{n_s \overline{n}_s}{(\ell_1\!+\!\ell_2)^2} 
 - i \frac{n_s \overline{n}_s}{(\ell_1\!+\!\ell_2)^2}
 + i \frac{n_t \overline{n}_t}{(\ell_2\!+\!\ell_3)^2} & = 0 \,,
\label{cancellation4}
\eeal
where we used \eqns{dilatonaxionexample}{twoscalarexample} to obtain zero.

Cancellations also happen among cuts where some of internal states are gravitons and some are matter states, that is, configurations such as $M_4^{\rm tree} (l^{++}_{1 h}\!,l^{--}_{2 h}\!, \ell^{-+}_{3\,\varphi}\!,\ell^{+-}_{4\,\varphi})$ for the middle blob. These are again effectively the same type of cancelation that has to happen for one-loop amplitudes. In summary, these cancelations leave the full four-dimensional cut corresponding to \fig{twoloopcut4pt} completely free of dilaton and axion contributions.  This shows that our prescription~\eqref{BCJformGravity} for calculating pure gravity amplitudes works correctly for this two-loop four-point example. We have repeated the same calculation \eqref{cancellation4} in the various supersymmetric settings listed in \tab{DCconstructions}.
In all cases, the unwanted matter
corresponding to $X$ and $\overline{X}$ multiplets cancel out.

\begin{figure}[t]
      \centering
      \includegraphics[scale=1.00]{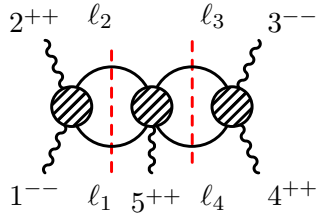}
      \vspace{-5pt}
\caption[a]{\small A two-loop cut of the five-graviton amplitude}
\label{twoloopcut5pt}
\end{figure}

The four-point analysis can be repeated in exactly the same way for the two-loop cut
of the five-graviton amplitude $ M(1^{--}_h\!,2^{++}_h\!,3^{--}_h\!,4^{++}_h\!,5^{++}_h) $.
Its most nontrivial cut is shown in \fig{twoloopcut5pt}.
To analyze it, we can use the tree-level input of \sect{tree5pt}.
At five points, the cancellation analogous to \eqn{cancellation4} takes the explicit form,
\small
\beal
     M_5^{\rm tree} (\ell^{-+}_{1\,\varphi}\!,\ell^{+-}_{2\,\varphi}\!,
                     \ell^{-+}_{3\,\varphi}\!,\ell^{+-}_{4\,\varphi}\!,5^{++}_h)
 - 2 M_5^{\rm tree} (\ell^{-+}_{1\,\varphi}\!,\ell^{+-}_{2\,\varphi}\!,
                     \ell^{-+}_{3\, \varphi'},\ell^{+-}_{4\, \varphi'},5^{++}_h) 
 - i \sum_{i=1}^5    \frac{n_i \overline{n}_i}{D_i}
 + i \sum_{i=6}^{10} \frac{n_i \overline{n}_i}{D_i} & = 0 \,,
\label{cancellation5}
\eeal
\normalsize
which is zero due to \eqns{FDC5}{FDC5mixed}.
We have also checked numerically that a similar cancellation occurs for cuts of the same topology with up to three more gravitons in the outer two blobs
and up to two more gravitons in the middle blob of this cut topology.

It is interesting to note that if we try to perform similar exercises and cut checks for double copies of fundamental scalars, instead of fermions, we encounter trouble.  Although the resulting ghost contributions will cancel the dilaton and axion in one-loop amplitudes, at two loops the construction fails to produce pure theories. This happens already for the cut analogous to \fig{twoloopcut4pt}, with fermion double copies replaced by scalar ones. In the next section, we give an indirect argument that confirms that scalar double copies are unsuitable for canceling the dilaton and axion.

\subsection{General multiloop argument}
\label{multiloop}

In this section, we argue that our prescription cancels the dilaton and axion from four-dimensional unitarity cuts at any loop order. We expect the construction to also produce correct $(4-2\epsilon)$-dimensional unitarity cuts, but since such cuts can be very subtle and regularization dependent, we give no explicit argument in that case.

For simplicity, we restrict the argument to Einstein gravity as the supersymmetric cases should be straightforward generalizations thereof. We want to show that a generic four-dimensional unitarity cut,
\be
   \sum_{S~{\rm states}} M^{\tree}_{(1)} M^{\tree}_{(2)}
                   \dots M^{\tree}_{(k)} \,,
\label{cuttrees}
\ee
constructed from double copies of YM diagrams according to our prescription, contains only graviton internal states.

In the bosonic case, our prescription involves the product
of two copies of (YM + fund. fermion) theories, where the states are tensored according to their gauge-group representation.
Let us use the fact that at tree level the states and partial amplitudes in this YM theory can be mapped to those of ${\cal N}=1$ SYM simply by replacing the fundamental-representation color factors with those in the adjoint (cf. \eqn{representationswap}).
Indeed, from the works of refs.~\cite{Drummond:2008cr,Dixon:2010ik,Reuschle:2013qna,Melia:2013bta,Melia:2013epa}, we know that the color-stripped tree-level amplitudes of ${\cal N}=1$ SYM (which are a subsector of ${\cal N}=4$ SYM)
can be combined to construct all the tree amplitudes in massless multiple-flavor QCD, and vice versa.
Therefore, for convenience, we map the states of (YM + fund. fermion)-theory
to a subsector of ${\cal N}=4$ SYM, and thus we can label the fields by the familiar SU(4) R-charge indices.

Let us look at the on-shell gravitational states
that the tensor product of two copies of (YM + fund. fermion)-theory gives us
according to the prescription~\eqref{BCJformGravity}.
If we reserve the R-charge indices $1,\dots,4$ for the left side of the double copy
and $5,\dots,8$ for the right side,
then the we have tensor products of the gluons $(A^{+} \oplus A^{-}_{1234}) \otimes (A^{+} \oplus A^{-}_{5678})$,
as well as of the matter states $\psi^{+}_1 \otimes \psi^{-}_{678}$
and $\psi^{-}_{234} \otimes \psi^{+}_{5}$, where the latter should be promoted to ghosts inside loop amplitudes.
This results in the following spectrum:
\be
   S=\{h^{+},~\varphi_{1234},~\varphi_{5678},~
         \hat{\varphi}_{1\,678},~\hat{\varphi}_{234\,5},~h^{-}_{1234\,5678}\} \,,
\ee
where the hat notation marks the states that are treated as ghosts in the loop amplitudes.

Now let us consider the simplest possible R-charge rotation,
a permutation of the indices ${\cal R}_{25}=\{2 \leftrightarrow 5\}$,
which gives the following set of states:
\be
   S'=\{h^{+},~\varphi_{134\,5}, ~\varphi_{2\,678},~
          \hat{\varphi}_{1\,678},~\hat{\varphi}_{234\,5},~h^{-}_{1234\,5678}\} \,.
\ee
Although we will not prove it, we claim that the gravitational tree amplitudes, and thus the unitarity cut \eqref{cuttrees}, following from a double-copy construction should be invariant under this rotation. This can be argued because of the close relationship between the tree-level kinematical numerators of (YM + fund. fermion)-theory and those of ${\cal N}=1$ SYM,
which in turn can be obtained from ${\cal N}=4$ SYM.
If this claim is true, we can observe
that the rotated spectrum $S'$ corresponds to double copies $A^{+} \otimes A^{+}$,
$A^{-}_{1234} \otimes A^{-}_{5678}$,
$(\hat{\psi}^{+}_1 \oplus \psi^{+}_2) \otimes \psi^{-}_{678}$ and
$(\hat{\psi}^{-}_{234} \oplus \psi^{-}_{134}) \otimes \psi^{+}_{5}$.
This can be considered to belong to a color-kinematics construction
with adjoint and fundamental fields,
where on one side we have (YM + fund. fermion + fund. fermion ghost)
and on the other side (YM + fund. fermion).
If the fundamental fields are restricted to live on internal lines of the loop amplitudes,
as they are in our construction,
then the amplitudes in the (YM + fund. fermion + fund. fermion ghost)-theory receives no contributions from the fermions, because they must cancel each other entirely.
Since the numerators corresponding to fundamental particles
vanish on one side of the double copy,
the gravity amplitude receives no contributions from fermion double copies. 
We can then remove the scalars produced by these double copies
and conclude that the spectrum~$S'$ is secretly equivalent to the pure gravity spectrum
\be
   S''=\{h^{+},~h^{-}_{1234\,5678}\} \,.
\ee

In other words, in our prescription,
the four-dimensional unitarity cuts should be equivalent to those of the pure gravity theory
\beal
   \sum_{S~{\rm states}}   M^{\tree}_{(1)} M^{\tree}_{(2)}
                     \dots M^{\tree}_{(k)} & =
   \sum_{S'~{\rm states}}  M^{\tree}_{(1)} M^{\tree}_{(2)}
                     \dots M^{\tree}_{(k)} \\ & =
   \sum_{S''~{\rm states}} M^{\tree}_{(1)} M^{\tree}_{(2)}
                     \dots M^{\tree}_{(k)} \,.
\eeal
Indeed, this is consistent with the explicit calculations in \sect{cutchecks}.

Note that for the above argument to be valid,
something must fail when applied to scalar double copies,
since we observed that the calculations in \sect{cutchecks} do not work for them.
To see this, consider the possible scalar prescription that relies on tensor products
$(A^{+} \oplus A^{-}_{1234}) \otimes (A^{+} \oplus A^{-}_{5678})$ and
$\phi_{12} \otimes \phi_{78}$ and $\phi_{34} \otimes \phi_{56}$.
This gives the following states:
\be
   S_\phi=\{h^{+},~\varphi_{1234},~\varphi_{5678},~
           \hat{\varphi}_{12\,78},~\hat{\varphi}_{34\,56},~h^{-}_{1234\,5678}\} \,.
\ee
Now perform, for example, the rotation
${\cal R}_{15;26} = \{1 \leftrightarrow 7, 2\leftrightarrow 8 \}$
of the R-charge indices:
\be
   S'_\phi=\{h^{+},~\varphi_{34\,78},~\varphi_{12\,56},
             ~\hat{\varphi}_{12\,78},~\hat{\varphi}_{34\,56},~h^{-}_{1234\,5678}\} \,.
\ee
Assuming that this is a valid double-copy spectrum the resulting matter can be thought of either as
$\phi_{12} \otimes (\phi_{56} \oplus \hat{\phi}_{78})$ and
$\phi_{34} \otimes (\hat{\phi}_{56} \oplus \phi_{78})$, or as
$(\phi_{12} \oplus \hat{\phi}_{34}) \otimes \phi_{56}$ and
$(\hat{\phi}_{12} \oplus \phi_{34}) \otimes \phi_{78}$, or even as $(\hat{\phi}_{12} \oplus \phi_{34}) \otimes (\hat{\phi}_{56} \oplus \phi_{78})$.
The first two cases look similar to the above fermionic situation, so we can attempt to interpret either of them as coming from a color-kinematics construction
with a (YM + fund. scalar + fund. scalar ghost)-theory on one side of the double copy. However, one can check that there will be an irregular four-scalar interaction that couples the ghosts to the non-ghost scalars, preventing a complete cancellation of the matter. Needless to say, this cannot happen to fermions. The third factorization $(\hat{\phi}_{12} \oplus \phi_{34}) \otimes (\hat{\phi}_{56} \oplus \phi_{78})$ is an even stranger case, since either side appears to contain a theory with one complex scalar whose antiparticle is a ghost, which is clearly not sensible.

Of course, one can consider doing other swaps of the R-symmetry indices,
but all possible permutations have similar problems.
While this does not conclusively prove that the scalar double-copy prescription fails for pure gravities, it does show that the argument that worked for fermion double copies
encounter obstructions when applied to scalars. In any case, the explicit cut checks in the previous section show that starting at two loops the scalar double copy fails to properly cancel the dilation and axion states.

\section{One-loop four-point gauge-theory amplitudes with internal matter}
\label{4ptYMsection}

In this section, we work out the duality-satisfying numerators
for massless one-loop four-point amplitudes with four external vector-multiplet legs and internal fundamental matter running in the loop.

\subsection{Reducing numerators to a master}
\label{generalapproach}

\begin{figure}[t]
      \centering
      \includegraphics[scale=0.80]{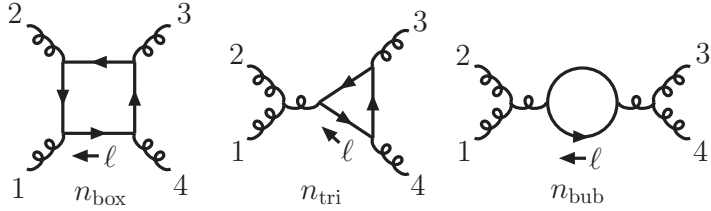}
      \vspace{-5pt}
\caption[a]{\small The three graphs that contribute to one-loop four-point amplitudes.
Curly lines correspond to adjoint vector multiplets
and solid lines represent fundamental matter
(or, in general, particles of any representation).
Our convention is that the external momenta are outgoing in all graphs.}
\label{ChiralBoxesFigure}
\end{figure}

\begin{figure}[t]
      \centering
      \includegraphics[scale=0.80]{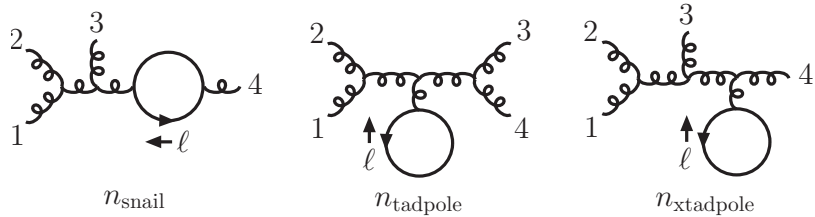}
      \vspace{-5pt}
\caption[a]{\small Three graphs that do not contribute to the amplitudes in dimensional regularization because they give scaleless integrals.}
\label{SnailTadpolesFigure}
\end{figure}

At four points, the master diagram that determines all other one-loop topologies is the box in \fig{ChiralBoxesFigure}, whose numerator we denote as
\be
   n_{\rm box}(1,2,3,4,\ell) \,,
\ee
where $\ell$ is the loop momentum and
$1,\dots,4$ are collective labels for the momentum, helicity and particle type of the external legs.
Other numerators can be obtained from the master using a set of kinematic  identities that follow from the color-kinematics duality.
Thanks to our conventions for the color algebra,
\eqns{jacobi}{commutation},
the dual kinematic relations for both fundamental and adjoint loop states translate to antisymmetrization of the external legs, which we denote by a commutator bracket.
In this way, the triangle and the bubble numerators, shown in \fig{ChiralBoxesFigure},  are obtained from the box as
\beal
   n_{\rm tri}(1,2,3,4,\ell) & \equiv n_{\rm box}([1,2],3,4,\ell) \,, \\ 
   n_{\rm bub}(1,2,3,4,\ell) & \equiv n_{\rm tri}(1,2,[3,4],\ell)
                                    = n_{\rm box}([1,2],[3,4],\ell) \,.
\label{Jac1}
\eeal
More explicitly,
$ n_{\rm tri}(1,2,3,4,\ell) = n_{\rm box}(1,2,3,4,\ell) - n_{\rm box}(2,1,3,4,\ell) $,
etc.
Further antisymmetrization gives the numerators of snail- and tadpole-type graphs,
shown in \fig{SnailTadpolesFigure}:
\beal
   n_{\rm snail}(1,2,3,4,\ell) & \equiv n_{\rm box}([[1,2],3],4,\ell) \,, \\  
   n_{\rm tadpole}(1,2,3,4,\ell) & \equiv n_{\rm box}([[1,2],[3,4]],\ell) \,, \\
   n_{\rm xtadpole}(1,2,3,4,\ell) & \equiv n_{\rm box}([[[1,2],3],4],\ell) \,.
\label{Jac2}
\eeal
Here a ``snail'' is synonymous to a bubble on an external leg,
whereas ``tadpole'' and ``xtadpole'' stand for tadpoles
on an internal leg and an external leg, respectively.
These three diagrams will integrate to zero in dimensional regularization,
as they will be identified with scaleless integrals. 

All graphs that we considered so far have the counterclockwise orientation
of the fundamental color flow. At one loop, a symmetry trick can be used to obtain the numerators with the internal-matter arrow reversed.
The matter-conjugated box diagram is given by mirroring the graph,
so that it is equivalent to the master box with all arguments reversed,
\be
   \overline{n}_{\rm box}(1,2,3,4,\ell) = n_{\rm box} (4,3,2,1,-\ell) \,,
\label{reverse}
\ee
and the others are obtained by conjugating \eqns{Jac1}{Jac2}.
In \sect{symmetriescuts}, we give an alternative definition of $\overline{n}_i$
using the building blocks of the numerators.

In addition to the above relations, we need to constrain the numerators
with the theory-specific physical information.
This is done using the unitarity method~\cite{Bern:1994zx,Bern:1994cg}.
\Eqns{Jac1}{Jac2} together with the unitarity cuts are equivalent
to imposing the color-kinematics duality through functional equations.
In the absence of a direct way to obtain the duality-satisfying numerators,
we will find a solution of these equations using an ansatz for the master box numerator.

\subsection{The ansatz construction}
\label{ansatzconstruction}

In this section, we explicitly construct a compact ansatz
for the master numerator $n_{\rm box}$,
suitable for a wide range of four-point amplitudes of various theories
(see alternative constructions
in refs.~\cite{Carrasco:2012ca,Nohle:2013bfa,Chiodaroli:2013upa,Ochirov:2013xba}).

To begin with, we list the elementary building blocks that are allowed to appear in the numerators. These include the Mandelstam invariants $s$, $t$ and $u$,
of which only two are independent,
as well as the scalar products $\ell \cdot k_i$ of the loop and external momenta.
Here we introduce a shorthand notation
for the three independent combinations of such products
that we use in the rest of the paper:
\be
      \ell_s \equiv 2\, \ell \cdot \! (k_1+k_2) \,, ~~~
      \ell_t \equiv 2\, \ell \cdot \! (k_2+k_3) \,, ~~~
      \ell_u \equiv 2\, \ell \cdot \! (k_1+k_3) \,.
\label{lproducts}
\ee

Another invariant is the Lorentz square of the loop momentum $\ell^2$.
However, in dimensional regularization, we must distinguish between
the four-dimensional square and the $D$-dimensional one.
We choose to work with the latter and the difference between the two:
\be
\mu^2 = \ell^2_{D=4} - \ell^2 \,,
\label{mu}
\ee
which corresponds to the square of the loop-momentum component
orthogonal to the four-dimensional spacetime
and is positive-valued in the metric signature $(+-\dots-)$.

Finally, we introduce the parity-odd Levi-Civita invariant
\be
      \eps(1,2,3,\ell)  \equiv {\rm Det}(k_1,k_2,k_3,\ell_{D=4}) \,,
\ee
which integrates to zero in gauge theory,
but do contribute to gravity when squared in the double-copy construction~\eqref{BCJformGravity}.

These building blocks already make it possible to write down an ansatz
for each of the distinct helicity configurations of the external states.
For four external gluons, a sufficient set of configurations consists of the split-helicity case $(1^-\!,2^-\!,3^+\!,4^+)$ and the alternating-helicity case $(1^-\!,2^+\!,3^-\!,4^+)$. All other external-state configurations in SYM are obtainable via supersymmetric Ward identities
\cite{Grisaru:1976vm,Grisaru:1977px,Parke:1985pn,Mangano:1990by} and permutations of labels. However, instead of using such identities we choose to rely the extended on-shell superspace formalism of \sect{supermultiplets}
to take care of the bookkeeping of external states.
Namely, for the four-point MHV amplitude, we introduce variables
\be
   \kappa_{ij} = \frac{ \spb{1}.{2} \spb{3}.{4} }{ \spa{1}.{2}\!\spa{3}.{4} }
   \delta^{(2{\cal N})}(Q) \spa{i}.{j}^{4-{\cal N}} \theta_{i} \theta_{j}  \,,
\label{kappa}
\ee
where the auxiliary parameter $\theta_{i}$ mark that the external leg $i$
belong to the anti-chiral vector multiplet, $\overline{V}_{\cal N}$, also containing the negative-helicity gluon. This implies that the unmarked legs are of the chiral type $V_{\cal N}$, which contains the positive-helicity gluon.
The super momentum delta function $\delta^{(2{\cal N})}(Q)$ ensures that the
${\cal N}$-fold supersymmetry Ward identities are respected among the components.
The remaining kinematic factors guarantee that the gluonic components take on a simple familiar form
\be
   \kappa_{ij} = istA_4^{\rm tree}(\dots i^- \!\!\!\dots j^- \!\!\!\dots)
                 (\eta_i^1 \eta_i^2 \eta_i^3 \eta_i^4)
                 (\eta_j^1 \eta_j^2 \eta_j^3 \eta_j^4) +\ldots \,.
\label{kappagluon}
\ee
For example, the pure-gluon component of $\kappa_{12}$ is explicitly  
$\spa{1}.{2}^2 \spb{3}.{4}^2$ using the spinor-helicity variables.
Thus, for a given graph topology we can ignore the helicity labels and instead work with a single generating function, spanned by the six independent $\kappa_{ij}$
that encode all the component numerators.

Now we can write down an ansatz for the master numerator of a superamplitude with external vector multiplets and any type of internal massless particles,
\be
   n_{\rm box}(1,2,3,4,\ell) = \sum_{1 \le i<j \le 4}
      \frac{\kappa_{ij}}{s_{ij}^{N}}
      \Big( \sum_{k} a_{ij;k} M_{k}^{(N)}
          + i \eps(1,2,3,\ell) \sum_{k} \tilde a_{ij;k} M_{k}^{(N-2)}
      \Big) \,,
\label{ansatz}
\ee
where $a_{ij;k}$ and $\tilde a_{ij;k}$ are free parameters of the ansatz,
$s_{ij} \equiv (k_i+k_j)^2$ are the Mandelstam invariants,
and $M^{(N)}_k$ denotes a monomial of dimension $2N$
built out of $N$ products of quadratic Lorentz invariants.
The monomials are drawn from the following set:
\be
   M^{(N)} = \Big\{ \prod_{i=1}^N m_i ~ \Big|~
                    m_i \in \{ s,\,t,\,\ell_s,\,\ell_t,\,\ell_u,\,\ell^2,\,\mu^2 \} 
             \Big\} \,,
\label{monomialset}
\ee
which contains $C_{N+6}^6$ distinct elements,  $C_{n}^k$ being the binomial numbers.
Hence the ansatz have
$ 6 (C_{N+6}^6 + C_{N+4}^6) $ free coefficients in total.
We vary the parameter $N$ in \eqn{ansatz} depending on
the effective number of supersymmetries ${\cal N}_{\rm eff}$ of the amplitude, such that
\be
   N=4-{\cal N}_{\rm eff} \,,
\ee
which is in accord with the classic loop-momentum power-counting argument of ref.~\cite{Bern:1994cg} and the ansatz rules for color-kinematics numerators of ref.~\cite{Bern:2012uf}.

As a simple and illuminating example that
the ansatz \eqref{ansatz} is finely tuned to the correct answer,
we can consider the case of ${\cal N}=3$ SYM,
which is equivalent to ${\cal N}_{\rm eff}=4$ SYM on shell.
Thus, using $N=0$, the ansatz becomes the six-fold parametrization
\be
   n_{\rm box}^{{\cal N}=3}(1,2,3,4,\ell)=\sum_{1\le i<j\le4}a_{ij;0}\,\kappa_{ij} \,,
\ee
where $\kappa_{ij}$ is well defined using \eqn{kappa} for ${\cal N}=3$.
As is easily checked, for $a_{ij;0}=1$ the sum collapses to the known answer
proportional to the tree superamplitude:
\be
   n_{\rm box}^{{\cal N}=3} = n_{\rm box}^{{\cal N}=4}
                            = i s t A^{\rm tree}_4(k_i, \eta^A_i) =  \frac{ \spb{1}.{2} \spb{3}.{4} }{ \spa{1}.{2}\!\spa{3}.{4} } \delta^{(8)}(Q) \,.
\label{N4Box}
\ee

Finally, note that the ansatz \eqref{ansatz} was chosen so that
the numerators with minus-helicity legs $i$ and $j$ will only contain
poles in the $s_{ij}$ channel.
This is a nontrivial highly-constraining aesthetic condition.
We found that such pole structure gives a necessary and sufficient class
of non-localities for the four-point one-loop numerators,
if they are built as rational functions of gauge-invariant building blocks.
Certainly, numerator non-localities can be completely avoided
by using gauge-dependent building blocks,
such as formal polarization vectors~\cite{Bern:2013yya, Nohle:2013bfa},
but we choose to work with the former type of ansatz.

\subsection{Imposing symmetries and cuts}
\label{symmetriescuts}

Before proceeding with the calculations of specific theories, we impose some extra universal constraints on the box numerator ansatz. They are not always necessary for the color-kinematics duality to be satisfied, but they often simplify the construction of the amplitudes both in gauge theory and in gravity.

\begin{figure}
      \centering
      \includegraphics[scale=1.00]{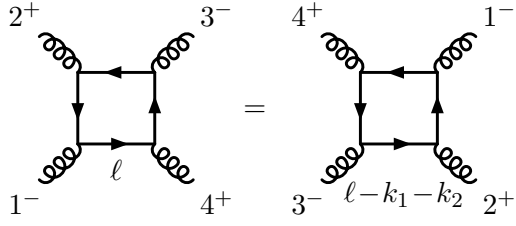}
\caption[a]{\small The cyclic symmetry of the alternating-helicity box diagram, which should be imposed on the numerator ansatz.}
\label{rotation}
\end{figure}

We choose to impose the cyclic symmetry natural to the box topology:
\be
   n_{\rm box}(1,2,3,4,\ell) = n_{\rm box}(2,3,4,1,\ell-k_1) \,. 
\label{rotatesymmetry}
\ee
Without this constraint the relabeling symmetry of the box numerator, and the descendent numerators \eqref{Jac1} and \eqref{Jac2}, would cease to be manifest.
In general, the constraint \eqref{rotatesymmetry} give identifications of various components in the supersymmetry expansion, and in particular it implies a two-site permutation symmetry of the alternating-helicity component $n_{\rm box}(1^-,2^+,3^-,4^+,\ell)$, shown in \fig{rotation}.

\begin{figure}[t]
      \begin{subfigure}{0.5\textwidth}
      \centering
      \includegraphics[scale=1.00]{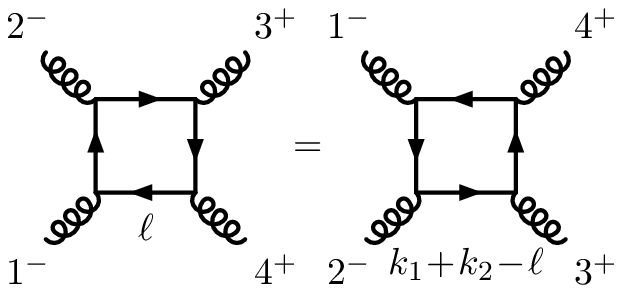}
      \caption[a]{\small \label{flip1}}
      \end{subfigure}
      \begin{subfigure}{0.5\textwidth}
      \centering
      \includegraphics[scale=1.00]{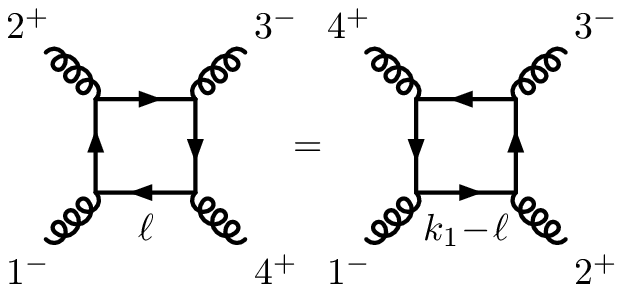}
      \caption[a]{\small \label{flip2}}
      \end{subfigure}
      \vspace{-5pt}
\caption[a]{\small The anti-fundamental boxes can be obtained
as the flipped fundamental boxes.
Through the matter-conjugation operation,
these relations impose constraints on the ansatz.}
\label{boxflip}
\end{figure}

Furthermore, we would like to extend the cyclic symmetry of the box numerator to a dihedral symmetry. The diagram flip relation \eqn{reverse} should then be imposed as a symmetry constraint on the ansatz. Recall that this equation reads
\be
   \overline{n}_{\rm box}(1,2,3,4,\ell) = n_{\rm box}(4,3,2,1,-\ell) \,.
\label{reverse2}  
\ee
The graph-symmetry origin of this equation is illustrated in \fig{boxflip}, where it is applied to the two distinct helicity components of the box numerator.
To use the above equation as a constraint, we need to precisely define how reversing the internal matter arrow operates on the ansatz \eqref{ansatz}. The matter conjugation should flip the chirality of the internal-loop matter particles while keeping the external chiral vectors unaltered. By using CPT invariance, we can alternatively regard this as a flip of the chirality of the external vector particles while keeping the internal matter unaltered. For consistent interactions, we also need to flip the sign of parity-odd momentum invariants.
The conjugated one-loop four-point MHV numerators are then
\be
   \overline{n}_i(1,2,3,4,\ell) = n_i(1,2,3,4,\ell)
      \big|_{ \kappa_{ij} \rightarrow \kappa^{\complement}_{ij} ,\,
              \eps(1,2,3,\ell)\rightarrow -\eps(1,2,3,\ell)} \,,
\label{matterConjugation}
\ee
where $ \kappa^{\complement}_{ij} = \kappa_{kl} $ marks the pair of legs
$\{k,l\}=\{1,2,3,4\} \setminus \{i,j\}$ unmarked by $\kappa_{ij}$. Combining \eqns{matterConjugation}{reverse2} give the desired extra constraint.

\begin{figure}[t]
      \centering
      \begin{subfigure}{0.32\textwidth}
            \includegraphics[scale=1.00]{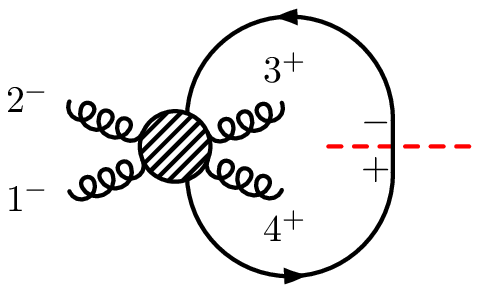}
      \caption[a]{\small \label{singlecut1}}
      \end{subfigure}
      \begin{subfigure}{0.32\textwidth}
            \includegraphics[scale=1.00]{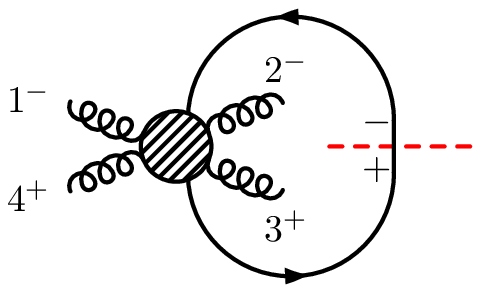}
      \caption[a]{\small \label{singlecut2}}
      \end{subfigure}
      \begin{subfigure}{0.32\textwidth}
            \includegraphics[scale=1.00]{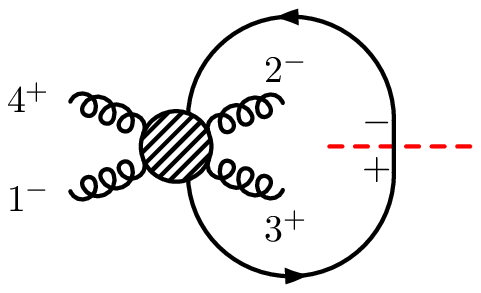}
      \caption[a]{\small \label{singlecut3}}
      \end{subfigure}
      \vspace{-5pt}
\caption[a]{\small Non-planar single-line cuts used to compute the numerators and amplitudes.}
\label{singlecuts}
\end{figure}

The ansatz~\eqref{ansatz} is ultimately constrained by the unitarity cuts.
We choose to work with the non-planar single-line cuts shown in \fig{singlecuts}. These cuts can be constructed by taking color-ordered six-point tree-level amplitudes
and identifying a conjugate pair of fundamental particles on opposite-site external legs. Because the tree amplitude is color-ordered, the identification of momenta on opposite ends of the amplitude does not produce singularities corresponding to soft or collinear poles.

As the internal loop momenta are subject to only one constraint, $\ell^2=0$,
the single-line cuts are sensitive to most terms in the integrand of the amplitude. Undetected terms correspond to tadpoles and snails (external bubbles), which invariably integrate to zero in a massless theory. Indeed, the tadpole and snail graphs previously described do not contribute to this cut, instead these will be indirectly constrained through the kinematic relations of the numerators. 

Even after the symmetries and unitarity cuts are imposed on the numerators,
there remain free parameters in the ansatz that can be interpreted as the residual generalized gauge freedom of the given amplitude representation.
For simplicity, we will fix this freedom by suitable aesthetic means,
as discussed in the next section.

\subsection{The amplitude assembly}
\label{amplitudeconstruction}

In this section, we provide the precise details on how to assemble the full one-loop gauge-theory amplitude from the numerators and the color factors. This discussion will be valid for all of the YM theories and amplitudes discussed in the subsequent sections.

The complete MHV (super-)amplitude can be written as
\be
   {\cal A}^{\text{1-loop}}_4 = \sum_{S_4}
      \Big(
            \frac{1}{8}\, {\cal I}_{\rm box}
          + \frac{1}{4}\, {\cal I}_{\rm tri}
          + \frac{1}{16}\,{\cal I}_{\rm bub}
      \Big) \,,
\label{ampl}
\ee
where the three (tensor) integrals correspond to the canonically labeled graphs
in \fig{ChiralBoxesFigure}, and the sum runs over the set $S_4$ that denotes the 24 permutations of the four external legs.
The rational numbers correspond to the symmetry factors that remove overcounting
both from the permutation sum and the internal phase-space integration.
In the case of fundamental matter multiplets in the loop,
each integral is formed by combining two diagrams with opposite orientations of the internal-matter arrow:
\be
   {\cal I}_i = \int\!\!\frac{d^{D}\ell}{(2\pi)^{D}}
      \frac{(c_i n_i + \overline{c}_i \overline{n}_i)}{D_i} \,,
\label{Ifund}
\ee
where $D=4-2\epsilon$ is the spacetime dimension in dimensional regularization,
and the denominators $D_i$ are products of the squared momenta
of the four internal lines in each graph in \fig{ChiralBoxesFigure},
thus accounting for the four propagators.
The color factors for the canonically labeled box, triangle and bubble graphs are
\beal
   c_{\rm box} & = \Tr(T^{a_1}T^{a_2}T^{a_3}T^{a_4}) \,, \\
   c_{\rm tri} \;\: & = \Tr([T^{a_1},T^{a_2}]T^{a_3}T^{a_4}) \,, \\
   c_{\rm bub} & = \Tr([T^{a_1},T^{a_2}][T^{a_3},T^{a_4}]) \,.
\label{fundColorFactors}
\eeal
The antifundamental color factors $\overline{c}_i$ are obtained from $c_i$ by replacing $T^a_{i\bar \jmath} \rightarrow T^a_{\bar \imath j}$ as discussed in \sect{colorkinematics}.

From \eqn{fundColorFactors}, it is obvious that the fundamental matter amplitude
can be rewritten in terms of only single traces,
thus giving color-ordered partial amplitudes in its decomposition
\be
   {\cal A}^{\text{1-loop,fund}}_4(1,2,3,4)
    = \!\!\sum_{\{i_1,i_2,i_3,i_4\} \in S_4/Z_4 \hspace{-30pt}}\!\!\!\!
      \Tr(T^{a_{i_1}}T^{a_{i_2}}T^{a_{i_3}}T^{a_{i_4}}\!)\,
      A^{\rm fund}_4(i_1,i_2,i_3,i_4) \,,
\label{fundOrdering}
\ee
where the sum runs over the cyclically inequivalent subset of $S_4$.

As explained in \sect{colorkinematics}, the fully adjoint amplitude can be obtained
from the amplitude in \eqn{ampl} after swapping the generators inside the color factors~\eqref{representationswap}.
The integrals ${\cal I}_i$ then take the following form: 
\be
   {\cal I}_i = \int\!\!\frac{d^{D}\ell}{(2\pi)^{D}}
      \frac{c_i^{\rm adj} n_i^{\rm adj}}{D_i} \,,
\label{Iadj}
\ee
where the adjoint color factors are given by
\beal
   c_{\rm box}^{\rm adj} & = \tf^{ba_1c}\tf^{ca_2d}\tf^{da_3e}\tf^{ea_4b} \,, \\
   c_{\rm tri}^{\rm adj} \: & = \tf^{a_1a_2c}\tf^{bcd}\tf^{da_3e}\tf^{ea_4b} \,, \\
   c_{\rm bub}^{\rm adj} & = \tf^{a_1a_2c}\tf^{bcd}\tf^{deb}\tf^{ea_3a_4} \,.
\label{adjColorFactors}
\eeal

In the standard SU($N_c$)-decomposition of the amplitude,
the adjoint color factors~\eqref{adjColorFactors} produce double traces:
\beal
   {\cal A}^{\text{1-loop,adj}}_4(1,2,3,4) =\;
    & N_c\!\!\!\! \sum_{\{i_1,i_2,i_3,i_4\} \in S_4/Z_4 \hspace{-30pt}}\!\!\!\!\!\;
      \Tr(T^{a_{i_1}}T^{a_{i_2}}T^{a_{i_3}}T^{a_{i_4}}\!)\,
      A^{\rm adj}_4(i_1,i_2,i_3,i_4) \\
    & + \!\! \!\! \sum_{\{i_1,i_2,i_3,i_4\} \in S_4/S_{4;2}\hspace{-33pt}}\!\!\!\!
      \Tr(T^{a_{i_1}}T^{a_{i_2}}\!)\Tr(T^{a_{i_3}}T^{a_{i_4}}\!)\,
      A^{\rm adj}_{4;2}(i_1,i_2,i_3,i_4) \,,
\label{adjOrdering}
\eeal
where $S_{4;2}$ is the subset of $S_4$ that leaves the double trace structure invariant.
The partial amplitudes $A^{\rm adj}_{4;2}$ in the subleading-color part are not independent but
related to the leading-color partial amplitudes $A^{\rm adj}_{4}$~\cite{Bern:1990ux,Bern:1994zx}.

A difference in  the structure of the one-loop color-ordered partial amplitudes
$A^{\rm fund}_{m}$ and $A^{\rm adj}_{m}$ is that the latter satisfy refection relations,
\be
      A^{\rm adj}_m(1,2,3,\dots,m) = (-1)^m A^{\rm adj}_m(m,\dots,,3,2,1) \,,
\label{mirror}
\ee
whereas the former generally do not. However, for the cases where the matter is effectively non-chiral, and thus $n_i=\overline{n}_i$, the reflection relation is restored, and moreover the partial amplitudes $A^{\rm adj}_4$ and $A^{\rm fund}_4$ in \eqns{fundOrdering}{adjOrdering} are identical.

\subsection{The amplitude with ${\cal N}=2$ fundamental or adjoint matter}
\label{N2}

Here we construct the four-point one-loop amplitude with a ${\cal N}=2$ fundamental matter multiplet $\Phi_{{\cal N}=2}$ circulating in the loop
and external adjoint vector multiplets ${\cal V}_{{\cal N}=2}$.
Because $\Phi_{{\cal N}=2}$ has the maximal amount of supersymmetry for matter, it is effectively a non-chiral multiplet. Only the color representation distinguishes it from its conjugate multiplet. Similarly, the color-stripped partial amplitudes are insensitive to the representation. To see this, note that the particle content of the multiplets $\Phi_{{\cal N}=2}$,
$\overline{\Phi}_{{\cal N}=2}$ in \eqn{chiralMatterMult}
and $\Phi_{{\cal N}=2}^{\rm adj}$ in \eqn{adjointNeq2} coincide
up to the gauge-group representation indices.
Using this fact, we can equate the one-loop kinematical numerators for different representations,
\be
      n_i^{{\cal N}=2,{\rm fund}} = \overline{n}_i^{{\cal N}=2,{\rm fund}} = 
      n_i^{{\cal N}=2,{\rm adj}} \,,
\label{fundadjN2}
\ee
which reduces the fundamental matter computation to the one with the adjoint matter.

Color-kinematics dual numerators for the component amplitude with external gluons
and adjoint ${\cal N}=2$ internal matter have already been constructed
in refs.~\cite{Carrasco:2012ca,Bern:2013yya,Nohle:2013bfa,Ochirov:2013xba}.
The result presented here is a minor variation of these known forms.
The main novelties are that we use the compact ansatz~\eqn{ansatz},
embed the external gluons into their vector supermultiplets,
generalize the internal color representation
and give the complete results for the integrand, including the $\mu$-terms and the snail diagrams.

Using the ansatz in \eqn{ansatz} for $N=4-{\cal N}_{\rm eff}=2$,
we have 174 free parameters to solve for.
By combining the equation
\be
      n_{\rm box}=\overline{n}_{\rm box} \,,
\label{CPTN2}
\ee
with the definition~\eqref{matterConjugation} of the antifundamental numerators,
we can reduce the ansatz by half, leaving only 87 free parameters.

Imposing the dihedral symmetries from~\sect{symmetriescuts} gives further constraints:
the cyclic symmetry~\eqref{rotatesymmetry} fixes 58 parameters.
A single further variable is fixed by the flip relation~\eqref{reverse},
which becomes a symmetry after imposing \eqn{CPTN2}. Enforcing the $D$-dimensional unitarity cuts in \fig{singlecuts} on the cut integrand fixes 24 out of 28 remaining parameters.
At this point, the cubic diagram representation satisfies all conditions
to be the correct amplitude, along with the built-in color-kinematics duality.

We choose to fix the remaining four parameters
by imposing aesthetic or practical constraints
to obtain compact and manageable numerator expressions.
Requiring that the snail numerator $n_{\rm snail}$
vanishes for any on-shell external momenta gives two additional conditions.
Finally, requiring that the parity-even part of the $s$-channel triangle
defined in \eqn{Jac1} is proportional to $s$ gives two more conditions.
The latter implicitly enforces no dependence on $\kappa_{12}$ and $\kappa_{34}$
in that triangle.\footnote{Alternatively, one can demand that the bubble numerator vanishes,
as was done in refs.~\cite{Carrasco:2012ca,Nohle:2013bfa,Ochirov:2013xba}.
This also fixes the last two parameters,
but gives a more complicated expression for the triangle.}

Having thus solved for all free parameters, we obtain the following box numerator:
\beal
      n_{\rm box}^{{\cal N}=2,{\rm fund}}
      = & \, (\kappa_{12}+\kappa_{34}) \frac{(s-\ell_s)^2}{2s^2}
        + (\kappa_{23}+\kappa_{14}) \frac{\ell_t^2}{2t^2}
        + (\kappa_{13}+\kappa_{24}) \frac{st+(s+\ell_u)^2}{2u^2} \\
      & + 2i \epsilon(1,2,3,\ell)\frac{\kappa_{13}-\kappa_{24}}{u^2}
        + \mu^2 \Big( \frac{\kappa_{12}+\kappa_{34}}{s}
                     +\frac{\kappa_{23}+\kappa_{14}}{t}
                     +\frac{\kappa_{13}+\kappa_{24}}{u} \Big) \,,
\label{N1evenBox}
\eeal
where the short-hand notation~\eqref{lproducts} for loop-momentum invariants is used and the parameters $\kappa_{ij}$ encoding the external multiplets
are defined in \eqn{kappa}.

The other numerators are given by the kinematic relations in \eqns{Jac1}{Jac2}.
To be explicit, the triangle numerator is
\beal
      n_{\rm tri}^{{\cal N}=2,{\rm fund}}
      = \,(\kappa_{23}+\kappa_{14})\frac{s(t-2\ell_t)}{2t^2}
        - (\kappa_{13}+\kappa_{24})\frac{s(u-2\ell_u)}{2u^2} & \\
       \null  + 2i \epsilon(1,2,3,\ell)
          \Big( \frac{\kappa_{23}-\kappa_{14}}{t^2}
              + \frac{\kappa_{13}-\kappa_{24}}{u^2} \Big) & \,,
\eeal
and the internal bubble numerator is
\be
      n_{\rm bub}^{{\cal N}=2,{\rm fund}}
      = s \Big( \frac{\kappa_{23}+\kappa_{14}}{t}-\frac{\kappa_{13}+\kappa_{24}}{u}\Big) \,.
\ee

In principle, we could ignore any further contributions to the integrand of the amplitude,
since snail (external bubble) and tadpole diagrams should integrate to zero
in dimensional regularization. 
However, for completeness, we give the result for the only non-vanishing diagram
of this type -- the snail.
Because of ${\cal N}=2$ supersymmetry power counting, it must include an overall factor $k_4^2$ (in analogy to a one-loop propagator correction), which vanishes on shell.
This contribution is not visible in our ansatz,
because the external legs were placed on shell from the very start.
Nevertheless, a careful analysis of a singular two-particle unitarity cut reveals
that the numerator of the snail diagram shown in \fig{SnailTadpolesFigure} is given by
\be
      n_{\rm snail}^{{\cal N}=2,{\rm fund}}
      = -\frac{k_4^2}{2} \Big( \frac{\kappa_{23}+\kappa_{14}}{t}
                             - \frac{\kappa_{13}+\kappa_{24}}{u} \Big) \,.
\ee
As the snail graph has a propagator $1/k_4^2$,
the above numerator gives a finite contribution to the integrand,
after the indeterminate form $0/0$ is properly canceled out.
Once this is done, the snail-diagram contribution can be included into the amplitude~\eqref{ampl}
as
\be
      \frac{1}{4} \sum_{S_4} {\cal I}_{\rm snail} \,,
\label{snailampl}
\ee
where ${\cal I}_{\rm snail}$ is defined by eqs.~\eqref{Ifund} or~\eqref{Iadj}
with the respective color factors
\be
      c_{\rm snail} = \Tr([[T^{a_1},T^{a_2}],T^{a_3}]T^{a_4})
      \qquad \text{or} \qquad
      c_{\rm snail}^{\rm adj} = \tf^{a_1a_2c}\tf^{ca_3d}\tf^{bde}\tf^{ea_4b} \,.
\label{snailFundColorFactor}
\ee

Of course, the snail diagram~\eqref{snailampl} still integrates to zero,
so it would be justified to drop it.
Nevertheless, we choose to explicitly display the snail graph
because there is some potential interest in the planar ${\cal N}=2$ integrand,
in analogy with the recent advances with the planar ${\cal N}=4$ integrand~\cite{ArkaniHamed:2010kv,ArkaniHamed:2012nw}.
Lastly, the two remaining numerators $n_\text{tadpole}$ and $n_\text{xtadpole}$ are
manifestly zero, consistent with naive expectations in ${\cal N}=2$ supersymmetric theories.

If we now assemble the full amplitude~\eqref{ampl}
and recast it into the color-ordered form~\eqref{fundOrdering},
we obtain the two inequivalent partial amplitudes in $D=4-2\epsilon$,
which are most easily expressed as
\begin{subequations} \begin{align}
      A^{{\cal N}=2,{\rm fund}}_4(1^-\!,2^-\!,3^+\!,4^+) =
            \frac{i \braket{12}^2 [34]^2}{ (4\pi)^{D/2} } 
            \bigg\{\!-&\frac{1}{st} I_2(t) \bigg\} \,, \\
      A^{{\cal N}=2,{\rm fund}}_4(1^-\!,2^+\!,3^-\!,4^+) =
            \frac{i \braket{13}^2 [24]^2}{ (4\pi)^{D/2} }
            \bigg\{\!-&\frac{r_{\Gamma}}{2u^2}
                       \left( \ln^2\!\left( \frac{-s}{-t} \right)\!+ \pi^2 \right)
                     + \frac{1}{su} I_2(s) + \frac{1}{tu} I_2(t)
            \bigg\} \,,
\end{align} \label{N2fund} \end{subequations}
\!\!where $I_2$ is the standard scalar bubble integral
\be
      I_2(t) = \frac{r_{\Gamma}}{\eps (1-2\eps)} (-t)^{-\eps} \,.
\label{I2}
\ee
Here and below, the integrated expressions are shown up to $O(\epsilon)$,
and the standard prefactor of dimensional regularization is
\begin{equation}
      r_{\Gamma} = \frac{ \Gamma(1+\epsilon) \Gamma^2(1-\epsilon) }
                        { \Gamma(1-2\epsilon) } \,.
\label{rgamma}
\end{equation}

Due to \eqn{fundadjN2}, the partial amplitudes~\eqref{N2fund} are the coefficients of both the clockwise and counterclockwise fundamental color traces in the color-dressed amplitude.
Moreover, they coincide with the primitive amplitudes $A^\text{1-loop,adj}_4$
with adjoint ${\cal N}=1$ matter in the loop (same as adjoint ${\cal N}=2$ in our convention, see \eqn{adjointNeq2}), which is the version that is best known in the literature~\cite{Bern:1994cg,Bern:1995db}.

\subsection{The amplitude with ${\cal N}=1$ fundamental matter}
\label{N1}

In this section, we work out the numerators of the four-point one-loop amplitude
with a ${\cal N}=1$ fundamental matter multiplet $\Phi_{{\cal N}=1}$
circulating in the loop and adjoint vectors ${\cal V}_{{\cal N}=1}$
on the external legs.
The result is the first known color-kinematics representation of this amplitude.

At one loop, the amplitude, along with its numerators,
can be naturally decomposed into two simpler components:
the parity-even and parity-odd contributions:
\begin{subequations} \begin{align}
   n_{i}^{{\cal N}=1, {\rm fund}} & =
      \frac{1}{2}\, n_{i}^{{\cal N}=1, {\rm even}}
    + \frac{1}{2}\, n_{i}^{{\cal N}=1, {\rm odd}} \,,
\label{nfund} \\
   \overline{n}_{i}^{{\cal N}=1, {\rm fund}} & =
      \frac{1}{2}\, n_{i}^{{\cal N}=1, {\rm even}}
    - \frac{1}{2}\, n_{i}^{{\cal N}=1, {\rm odd}} \,,
\label{nantifund}
\end{align} \label{nfundboth} \end{subequations}
\!\!where we define
\begin{subequations} \begin{align}
      n_{i}^{{\cal N}=1, {\rm even}} & \equiv n_{i}^{{\cal N}=1, {\rm fund}}
                                     + \overline{n}_{i}^{{\cal N}=1, {\rm fund}} \,,
\label{neven} \\
      n_{i}^{{\cal N}=1, {\rm odd}}  & \equiv n_{i}^{{\cal N}=1, {\rm fund}}
                                     - \overline{n}_{i}^{{\cal N}=1, {\rm fund}} \,.
\label{nodd}
\end{align} \label{nevenodd} \end{subequations}
According to \eqn{nplusnbar}, summing over the fundamental and antifundamental one-loop numerators effectively corresponds to promoting the ${\cal N}=1$ multiplet to the adjoint representation, which by \eqn{adjointNeq2} is equivalent to a contribution of a ${\cal N}=2$ multiplet. Thus we have the following equalities:
\be
      n_{i}^{{\cal N}=1, {\rm even}} = n_{i}^{{\cal N}=2, {\rm adj}}
                                     = n_{i}^{{\cal N}=2, {\rm fund}} \,.
\ee

Since we have already found compact expressions for the ${\cal N}=2$ numerators
in \sect{N2}, we can now focus entirely on the parity-odd ${\cal N}=1$ numerators.
Unlike the parity-even ones, they have the expected amount of supersymmetry,
hence ${\cal N}_{\rm eff}=1$.
Using the ansatz~\eqref{ansatz} as before, except with $N=4-{\cal N}_{\rm eff}=3$,
gives us 546 free parameters to solve for.
Similarly to the procedure in the previous section,
we can immediately eliminate half of the parameters by imposing the defining property
that the box numerator should be odd under matter conjugation:
\be
      n_{\rm box}^{\rm odd}=-\overline{n}_{\rm box}^{\rm odd} \,,
\label{CPTN1}
\ee
which reduces the problem to 273 undetermined parameters.

Next comes the dihedral symmetry of \sect{symmetriescuts}:
the cyclic symmetry~\eqref{rotatesymmetry} fixes 208 parameters.
An additional 20 are constrained by the flip relation~\eqref{reverse},
which becomes a symmetry after imposing \eqn{CPTN1}. 

Out of the remaining 45 free parameters, 20 are fixed
by the four-dimensional unitarity cuts shown in \fig{singlecuts}.
Requiring that the snail numerator vanishes on shell gives 17 additional constraints. Demanding that the power counting of the numerator
is at worst $\ell^m$ for one-loop $m$-gons gives 3 more conditions,
leaving only five parameters to fix.
One can check that four of them correspond to genuine freedom of the color-kinematics representations,
and one parameter is determined by $D$-dimensional unitarity cuts.
However, since it is difficult to analytically continue four-dimensional chiral fermions
to $D>4$, computing the correct unitarity cut is challenging.
Instead, in our final representation,
we denote this unfixed parameter by $a$
and manually choose the remaining four parameters
to the values that give more compact expressions.

The box numerator for the ${\cal N}=1$ odd contribution is then given by
\beal
      n_{\rm box}^{{\cal N}=1, {\rm odd}} \!=
          & - (\kappa_{12}-\kappa_{34}) \frac{(\ell_s-s)^3}{2s^3}
            - (\kappa_{23}-\kappa_{14}) \frac{\ell_t^3}{2 t^3}
            - (\kappa_{13}-\kappa_{24}) \frac{1}{2}
              \Big( \frac{\ell_u^3}{u^3} \!+\! \frac{3s\ell_u^2}{u^3}
              \!-\! \frac{3s\ell_u}{u^2} \!+\! \frac{s}{u} \Big) \\
          & - 2i \epsilon(1,2,3,\ell)  (\kappa_{13}+\kappa_{24}) \frac{2\ell_u-u}{u^3}
            + a \mu^2 (\kappa_{13}-\kappa_{24})\frac{s-t}{u^2} \,,
\label{N1oddBox}
\eeal
and the triangle and bubble numerators can be obtained using \eqn{Jac1}.
They are
\begin{align}
   n_{\rm tri}^{{\cal N}=1, {\rm odd}} \!=
      & \; (\kappa_{23}-\kappa_{14}) s
          \Big( \frac{3\ell_t^2}{2t^3} - \frac{3\ell_t}{2t^2} + \frac{1}{2t} \Big)
        - (\kappa_{13}-\kappa_{24}) s
          \Big( \frac{3\ell_u^2}{2u^3} - \frac{3\ell_u}{2u^2} + \frac{1}{2u} \Big) \nn \\
      & - 2i \epsilon(1,2,3,\ell)
          \Big( (\kappa_{23}+\kappa_{14}) \frac{2\ell_t-t}{t^3}
              + (\kappa_{13}+\kappa_{24}) \frac{2\ell_u-u}{u^3} \Big) \\
      & - a \mu^2 (\kappa_{23}-\kappa_{14})\frac{s-u}{t^2}
        + a \mu^2 (\kappa_{13}-\kappa_{24})\frac{s-t}{u^2} \,, \nn \\
n_{\rm bub}^{{\cal N}=1, {\rm odd}} \!=
      & - (\kappa_{23}-\kappa_{14}) \frac{3s\ell_t}{t^2}
        + (\kappa_{13}-\kappa_{24}) \frac{3s\ell_u}{u^2}
        + 4i \epsilon(1,2,3,\ell) \Big( \frac{\kappa_{23}+\kappa_{14}}{t^2}
                                      + \frac{\kappa_{13}+\kappa_{24}}{u^2} \Big) \,.
\end{align}
As for the three remaining numerators
$n_{\rm snail}$, $n_{\rm tadpole}$ and $n_{\rm xtadpole}$, defined by \eqn{Jac2},
they are manifestly zero in this representation.

We note that the combinations of $n_i$ and $c_i$ that appear in \eqn{Ifund}
can be rewritten as follows:
\be
      n_i^{{\cal N}=1,{\rm fund}} c_i
    + \overline{n}_i^{{\cal N}=1,{\rm fund}} \overline{c}_i
    = \frac{1}{2}\, n_i^{{\cal N}=1,{\rm even}}(c_i+\overline{c}_i)
    + \frac{1}{2}\, n_i^{{\cal N}=1,{\rm odd}}(c_i-\overline{c}_i) \,.
\ee
This implies that we can write the following relation for the color-dressed amplitudes:
\be
      {\cal A}^{{\cal N}=1,{\rm fund}}_4
    = \frac{1}{2}\, {\cal A}^{{\cal N}=2,{\rm fund}}_4
    + \frac{1}{2}\, {\cal A}^{{\cal N}=1,{\rm odd}}_4 \,,
\label{AN1}
\ee
where the last ``amplitude'' is constructed from the parity-odd numerators
and color factors,
\be
      c_i^{\rm odd}=c_i-\overline{c}_i \,.
\ee
Although chiral gauge anomalies are not the topic of the current work, it is interesting to note that this contribution contains all the anomalies of the ${\cal N}=1$ amplitude.

The two inequivalent color-ordered amplitudes for the odd part must coincide with
those for chiral fermions (see \sect{N0ferm}).
These amplitudes were, for example, computed from four-dimensional unitarity cuts and locality conditions in ref.~\cite{Huang:2013vha};
\begin{subequations} \begin{align}
      A^{{\cal N}=1,{\rm odd}}_4(1^-\!,2^-\!,3^+\!,4^+) & = 0 \,, \\
      A^{{\cal N}=1,{\rm odd}}_4(1^-\!,2^+\!,3^-\!,4^+) & =
          - \frac{i r_{\Gamma} \braket{13}^2 [24]^2}{ (4\pi)^{D/2} }
            \bigg\{ \frac{s\!-\!t}{2u^3}
                    \left( \ln^2\!\left( \frac{-s}{-t} \right)\!+ \pi^2 \right)
                 +\!\frac{2}{u^2} \ln\!\left( \frac{-s}{-t} \right)
                 -  \frac{s\!-\!t}{2stu}
            \bigg\} \,.
    \label{N1odd} 
\end{align}\end{subequations}
\! After comparing to these results, the integration of the color-kinematics representation~\eqref{N1oddBox} fixes the last parameter to $a=3/2$.
Interestingly, the precise value of $a$ is irrelevant for the gravity amplitudes constructed in \sect{gravity},
as the parameter $a$ drops out after the double-copy amplitudes are integrated.

\subsection{The MHV amplitude with fundamental ${\cal N}=0$ scalar matter}
\label{N0sc}

In this section, we construct the MHV amplitude with the fundamental scalar
$\Phi^{\rm scalar}_{{\cal N}=0}$ circulating in the loop and external adjoint vectors
${\cal V}_{{\cal N}=0}$, i.e. gluons.
As in the ${\cal N}=2$ case, after ignoring the color representation the scalar matter is effectively CPT-invariant and thus the color-stripped amplitudes for $\Phi_{{\cal N}=0}^{\rm scalar}$,
$\overline{\Phi}_{{\cal N}=0}^{\rm scalar}$ and $\Phi_{{\cal N}=0}^{\rm adj\;scalar}$
are the same. Therefore, we can equate the numerators of the different multiplets:
\be
      n_i^{\rm scalar} = \overline{n}_i^{\rm scalar} = 
      n_i^{\rm adj\;scalar} \,.
\label{fundadjN0}
\ee
The color-kinematics representation of the amplitude with an adjoint scalar
has been previously constructed~\cite{Bern:2013yya,Nohle:2013bfa}
in a fully-covariant form (using formal polarization vectors).
The result presented here will be a helicity-based computation of this amplitude.
The main novelties are that we use the compact ansatz~\eqref{ansatz}
that relies on the four-dimensional notation for the external legs,
and that we trivially generalize the internal color representation.
Note that we only give the MHV amplitude, leaving out, for brevity,
the amplitudes exclusively present in non-supersymmetric theories:
the all-plus-helicity and one-minus-helicity amplitudes,
and their parity conjugates (see ref.~\cite{Nohle:2013bfa} for these).

The ansatz~\eqref{ansatz} with ${\cal N}_{\rm eff}=0$
gives a parametrization with 1428 variables.
Imposing
\be
      n_{\rm box} = \overline{n}_{\rm box} \,,
\label{CPTN0}
\ee
immediately reduces the undetermined parameters by a factor of two.
The cyclic symmetry~\eqref{rotatesymmetry} fixes
512 out of 714 remaining parameters,
and the flip relation~\eqref{reverse} eliminates 29 more.
The $D$-dimensional unitarity cuts in \fig{singlecuts} fix 118 parameters,
leaving 55 free.

Next we impose practical constraints.
Similarly to refs.~\cite{Bern:2013yya,Nohle:2013bfa},
we demand that the snail diagram in \fig{SnailTadpolesFigure}
gives a scaleless integral:
otherwise it will not necessarily integrate to zero.\footnote{Imposing
that the snail numerator vanish would be ideal,
but this is not consistent with the ansatz~\eqref{ansatz}
and the cuts in the ${\cal N}=0$ case.}
A scaleless snail integral is achieved
if its numerator is allowed to be a function of only one scalar product
between the external and internal momenta, namely $k_4 \cdot \ell$,
where $k_4$ is the external momentum directly entering the bubble subgraph of the snail.
This means that the integral is a function of only $k_4^2=0$,
so it must vanish in dimensional regularization.
This constraint fixes 25 parameters.

Continuing with aesthetic conditions, we require that the triangle numerator
has no dependence on $\kappa_{12}$ or $\kappa_{34}$, which gives 20 more conditions. Demanding that the power counting of each $m$-gon numerator is at most $\ell^m$
gives 5 more constraints.
The remaining 5 parameters are fixed manually to obtain more compact expressions for the numerators.

As a result, we obtain the following box numerator for the one-loop four-gluon amplitude 
with a scalar in the loop:
\beal
   n_{\rm box}^{\rm scalar} \! =
      & - (\kappa_{12}+\kappa_{34})
          \Big( \frac{\ell_s^4}{4 s^4} - \frac{\ell_s^2 (2 \ell^2 + 3 \ell_s)}{4 s^3}
              + \frac{2 \ell^2 \ell_s + \ell_s^2 - 2 \mu^4}{2 s^2}
              - \frac{2 \ell^2 - \ell_s + s}{4 s} \Big) \\
      & - (\kappa_{23}+\kappa_{14})
          \Big( \frac{\ell_t^4}{4 t^4}
              - \frac{\ell_t^2 (2 \ell^2 - \ell_s - \ell_u + t)}{4 t^3}
              - \frac{\mu^4}{t^2} \Big) \\
      & - (\kappa_{13}+\kappa_{24})
          \Big( \frac{\ell_u^3 (\ell_u + 3 s)}{4 u^4}
              - \frac{\ell_u( \ell_u (2 \ell^2 - \ell_s) - \ell_s^2 + \ell_t^2
                            + 4 s (\ell^2 + \ell_u + 2 \mu^2))}{4 u^3} \\
      & ~~~~~~~~~~~~~~~~~~~~~~~~~~~~~~~
              - \frac{\ell_s^2 - \ell_t^2 + 3 \ell_u^2 + 4 \ell^2 t
                     + 8 \mu^2  (\ell_u - s + \mu^2)}{8 u^2}
              - \frac{\ell_s - s}{4 u} \Big) \\ 
      & - 2i \epsilon(1,2,3,\ell) (\kappa_{13}-\kappa_{24})
          \frac{\ell_u^2 - u \ell_u - 2 \mu^2 u}{u^4} \,.
\label{N0scalarBox}
\eeal
The other numerators are given by the kinematic relations~\eqref{Jac1}
and~\eqref{Jac2}. Explicitly, the triangle numerator is
\beal
   n_{\rm tri}^{\rm scalar} \, =
      ~ & (\kappa_{23}+\kappa_{14})
          \Big( \frac{3\ell_t^3s}{4t^4}
              - \frac{\ell_t (\ell_u^2-\ell_s^2+\ell_t\ell_u+4s(\ell^2+\ell_t+2\mu^2))}
                     {4 t^3} \\ & ~~~~~~~~~~~~~~~~~~\:\,
              - \frac{\ell_s^2+\ell_t^2-\ell_u^2+8\ell_t\mu^2-4s(\ell^2+2\mu^2)}{8 t^2}
              + \frac{2\ell^2-\ell_s+s}{4t}
          \Big) \\
     -\, & (\kappa_{13}+\kappa_{24})
          \Big( \frac{3 \ell_u^3 s}{4u^4}
              - \frac{\ell_u (\ell_t^2-\ell_s^2+\ell_t\ell_u+4s(\ell^2+\ell_u+2\mu^2))}
                     {4 u^3} \\ & ~~~~~~~~~~~~~~~~~~\:\,
              - \frac{\ell_s^2+\ell_u^2-\ell_t^2+8\ell_u\mu^2-4s(\ell^2+2\mu^2)}
                     {8 u^2}
              + \frac{2\ell^2-\ell_s+s}{4u}
          \Big) \\
     -\, & 2i \epsilon(1,2,3,\ell)
          \Big( (\kappa_{23}-\kappa_{14}) \frac{\ell_t^2 - t \ell_t - 2 \mu^2 t}{t^4}
             +  (\kappa_{13}-\kappa_{24}) \frac{\ell_u^2 - u \ell_u - 2 \mu^2 u}{u^4}
          \Big) \,, \!\!\!\!\!\!\!\!\!\!
\eeal
and the bubble numerator is
\beal
   n_{\rm bub}^{\rm scalar} \, =
      & - (\kappa_{23}+\kappa_{14})
          \Big( \frac{2s\ell_t^2}{t^3}
              + \frac{4u\ell^2 - \ell_u^2 + \ell_t^2 + \ell_s^2 - 8 s \mu^2}{4t^2}
              + \frac{\ell_s-s}{2t} \Big) \\
      & + (\kappa_{13}+\kappa_{24})
          \Big( \frac{2s\ell_u^2}{u^3}
              + \frac{4t\ell^2 - \ell_t^2 + \ell_u^2 + \ell_s^2 - 8 s \mu^2}{4u^2}
              + \frac{\ell_s-s}{2u} \Big) \\
      & + 4i \epsilon(1,2,3,\ell)
          \Big( \ell_t \frac{\kappa_{23}-\kappa_{14}}{t^3}
              + \ell_u \frac{\kappa_{13}-\kappa_{24}}{u^3} \Big) \,.
\eeal
There are also nonvanishing numerators for external bubbles and
external and internal tadpoles that can be obtained from \eqn{Jac2},
but since these integrals integrate to zero we do not display them here.

Assembling the pieces of the ${\cal N}=0$ amplitude for a fundamental
or an adjoint scalar is done along the lines of \sect{amplitudeconstruction}.
In both cases, the primitive color-stripped amplitudes are the same,
only the color dressing will differ between the gauge-group representations.
The primitive color-stripped amplitudes for a single scalar contribution
in the loop are known~\cite{Bern:1995db} to be:
\begin{subequations} \begin{align}
      A^{{\cal N}=0,{\rm scalar}}_4(1^-\!,2^-\!,3^+\!,4^+) & =
            \frac{i \braket{12}^2 [34]^2}{ (4\pi)^{D/2} } 
            \bigg\{\!- \frac{1}{6st} I_2(t) - \frac{r_{\Gamma}}{9st} \bigg\} \,, \\
      A^{{\cal N}=0,{\rm scalar}}_4(1^-\!,2^+\!,3^-\!,4^+) & =
            \frac{i \braket{13}^2 [24]^2}{ (4\pi)^{D/2} }
            \bigg\{\!- \frac{r_{\Gamma} st}{2u^4}
                       \left( \ln^2\!\left( \frac{-s}{-t} \right)\!+ \pi^2 \right) \\ & ~~~
                     - \left( \frac{s\!-\!t}{2u^3} - \frac{1}{6su} \right) I_2(s)
                     - \left( \frac{t\!-\!s}{2u^3} - \frac{1}{6tu} \right) I_2(t)
                     + \frac{r_{\Gamma}}{2u^2} - \frac{r_{\Gamma}}{9st}
            \bigg\} \nn \,,
\end{align} \label{N0scalar} \end{subequations}
\!\!and indeed our construction agrees with these.

\subsection{The MHV amplitude with fundamental ${\cal N}=0$ fermion matter}
\label{N0ferm}

Here we present the MHV amplitude with a fundamental chiral fermion
$\Phi^{\rm fermion}_{{\cal N}=0}$ circulating in the loop
and external adjoint vectors ${\cal V}_{{\cal N}=0}$, i.e. gluons.
This amplitude is actually a simple linear combination of the three amplitudes
discussed above, so no extra work is needed.
Note that, as for the scalar case, we only give the MHV amplitude.

The basic identity that we use is that a chiral fermion is obtained after subtracting the scalar $\Phi_{{\cal N}=0}^{\rm scalar}$ from the $\Phi_{{\cal N}=1}$ multiplet.
For the one-loop amplitudes this implies that
\be
   {\cal A}^{\rm fermion}_4 = {\cal A}^{{\cal N}=1,{\rm fund}}_4
                            - {\cal A}^{\rm scalar}_4 \\  
                            = \frac{1}{2}\,{\cal A}^{{\cal N}=2,{\rm fund}}_4
                            + \frac{1}{2}\,{\cal A}^{{\cal N}=1,{\rm odd}}_4
                            - {\cal A}^{\rm scalar}_4 \,,
\ee
where we used \eqn{AN1}.
Hence we can define the fundamental fermion numerators as:
\begin{subequations} \begin{align}
   n_i^{\rm fermion} \equiv
        \frac{1}{2}\,n_i^{{\cal N}=2,{\rm fund}}
      + \frac{1}{2}\,n_i^{{\cal N}=1,{\rm odd}}
      - n_i^{\rm scalar} \,,
\label{nfermion} \\
   \overline{n}_i^{\rm fermion} \equiv
        \frac{1}{2}\,n_i^{{\cal N}=2,{\rm fund}}
      - \frac{1}{2}\,n_i^{{\cal N}=1,{\rm odd}}
      - n_i^{\rm scalar} \,.
\label{nbarfermion}
\end{align} \label{nfermionboth} \end{subequations}

Note that, just as in the ${\cal N}=1$ case, the parity-odd contributions
to the chiral fermion amplitude come entirely from the odd ${\cal N}=1$ sector.
Thus, interestingly, the chiral-anomalous part of the chiral fermion amplitude
effectively has ${\cal N}=1$ supersymmetry.

\section{One-loop four-point supergravity amplitudes}
\label{gravity}

In this section, we assemble the duality-satisfying numerators
into various one-loop four-point supergravity amplitudes.
Similar to \eqn{ampl}, we write these amplitudes as
\be
{\cal M}^{\text{1-loop}}_4 = \Big(\frac{\kappa}{2}\Big)^4 \sum_{S_4}
      \Big( \frac{1}{8}\, {\cal I}_{\rm box}
          + \frac{1}{4}\, {\cal I}_{\rm tri}
          + \frac{1}{16}\,{\cal I}_{\rm bub} \Big) \,,
\label{ampl2}
\ee
where the three integrals correspond to the canonically labeled graphs
in \fig{ChiralBoxesFigure}, and the rational prefactors compensate for overcount
in the sum over the set $S_4$ of 24 permutations of the four external legs, as well as overcount in the phase space of the bubble graph.

\subsection{Matter amplitudes}
\label{mattergravity}

Here we check the double-copy construction
for gravity amplitudes with external graviton multiplets $H_{\cal N}$\footnote{More
precisely, to get $H_{\cal N}$ multiplets,
the external dilaton-axion multiplets $X$, $\overline{X}$ should be projected out,
which is straightforward for external states.}
and internal matter of the types shown in the right column of \tab{DCconstructions}.

First of all, consider the following double copy of the ${\cal N}=2$ numerators
of \sect{N2}:
\be
   {\cal I}_i^{{\cal N}=2+2,{\rm matter}} = \int\!\!\frac{d^{D}\ell}{(2\pi)^{D}} \,
      \frac{n_i^{{\cal N}=2,{\rm fund}}\,\overline{n}_i^{{\cal N}=2,{\rm fund}}
        \!+ \overline{n}_i^{{\cal N}=2,{\rm fund}}\,n_i^{{\cal N}=2,{\rm fund}}}{D_i} \,,
\label{Neq2p2Integrals}
\ee
which is a precise implementation of the last line and right column of \tab{DCconstructions}.
Due to the effective CPT invariance of the ${\cal N}=2$ matter numerators, this fundamental-representation double copy gives exactly twice the known~\cite{Carrasco:2012ca,Bern:2013yya,Nohle:2013bfa,Ochirov:2013xba}
adjoint double copy,
\be
   {\cal I}_i^{{\cal N}=2+2,{\rm matter}} = 2 \, {\cal I}_i^{{\cal N}=4,{\rm matter}}=2\!\int\!\!\frac{d^{D}\ell}{(2\pi)^{D}}
                  \frac{(n_i^{{\cal N}=2,{\rm adj}})^2}{D_i} \,.
\label{Neq2p2adjIntegrals}
\ee
The adjoint-representation double copy corresponds to a single ${\cal V}_{{\cal N}=4}$ multiplet contribution in the loop. Up to this factor of two,
we recover the ${\cal N}=4$ matter amplitude,
first computed by Dunbar and Norridge in ref.~\cite{Dunbar:1994bn}:
\beal
   M^{{\cal N}=4,{\rm matter}}_4(1^{--}\!,2^{--}\!,3^{++}\!,4^{++})
    = \frac{i r_{\Gamma} \braket{12}^4 [34]^4}{ (4\pi)^{D/2} } \frac{1}{2 s^4}
      \bigg\{\!& - t u \left( \ln^2\!\left( \frac{-t}{-u} \right)\!+ \pi^2 \right) \\
               & + s(t\!-\!u) \ln\!\left( \frac{-t}{-u} \right)\!+ s^2
      \bigg\} \,,
\label{N4gravitymatter}
\eeal
and the other component amplitudes are given by supersymmetry Ward identities.

Using the same relation for the ${\cal N}=2$ matter numerators,
and additionally taking into account that
\be
   n_i^{{\cal N}=1,{\rm fund}} + \overline{n}_i^{{\cal N}=1,{\rm fund}}
                               = n_i^{{\cal N}=2,{\rm adj}}\,,
\label{N1even}
\ee
the fundamental-representation double copy
in the second-to-last line of \tab{DCconstructions} becomes
\be
   {\cal I}_i^{{\cal N}=2+1,{\rm matter}} = \int\!\!\frac{d^{D}\ell}{(2\pi)^{D}} \,
      \frac{n_i^{{\cal N}=2,{\rm fund}}\,\overline{n}_i^{{\cal N}=1,{\rm fund}}
        \!+ \overline{n}_i^{{\cal N}=2,{\rm fund}}\,n_i^{{\cal N}=1,{\rm fund}}}{D_i}={\cal I}_i^{{\cal N}=4,{\rm matter}} \,.
\label{Neq2p1Integrals}
\ee
So these contributions also reduce to the adjoint-representation double copy~\eqref{Neq2p2adjIntegrals}
of the amplitude~\eqref{N4gravitymatter},
though this time without the factor of two.

The discussed two entries of \tab{DCconstructions} can be regarded merely
as a new perspective
on the corresponding adjoint-representation double copies.
However, the remaining part of the right column of \tab{DCconstructions}
is genuinely tied to the novel fundamental construction,
and not trivially related to adjoint double copies.
For instance, the $({\cal N}\!=\!1)^2$ matter double copy successfully exploits\footnote{Recall that ${\cal N}=1$ (chiral) adjoint matter works differently
since it effectively has ${\cal N}=2$ supersymmetry
and thus produces $({\cal N}\!=\!2)^2 = ({\cal N}\!=\!4)$ matter in the double copy.}  the chiral structure, giving
\be
   {\cal I}_i^{{\cal N}=1+1,{\rm matter}} = \int\!\!\frac{d^{D}\ell}{(2\pi)^{D}} \,
      \frac{n_i^{{\cal N}=1,{\rm fund}}\,\overline{n}_i^{{\cal N}=1,{\rm fund}}
        \!+ \overline{n}_i^{{\cal N}=1,{\rm fund}}\,n_i^{{\cal N}=1,{\rm fund}}}{D_i}=2 \, {\cal I}_i^{{\cal N}=2,{\rm matter}}  \,,
\label{Neq1p1Integrals}
\ee
where the last identity indicates that this construction corresponds
to twice the contribution of a single non-chiral ${{\cal N}=2}$ matter multiplet.
The graviton component of this matter amplitude was computed
in ref.~\cite{Dunbar:1994bn} using non-chiral building blocks,
its integrated form is
\beal
   M^{{\cal N}=2,{\rm matter}}_4(1^{--}\!,2^{--}\!,3^{++}\!,4^{++})
    = \frac{i r_{\Gamma} \braket{12}^4 [34]^4}{ (4\pi)^{D/2} }
      \bigg\{\! - \frac{t^2 u^2}{2s^6}
                  \left( \ln^2\!\left( \frac{-t}{-u} \right)\!+ \pi^2 \right) & \\
                + \frac{(t-u)(t^2 + 8tu + u^2)}{12s^5}
                  \ln\!\left( \frac{-t}{-u} \right)
                + \frac{t^2 + 14tu + u^2}{24s^4} &
      \bigg\} \,.
\label{N2gravitymatter}
\eeal
The fact that our chiral double copy agrees with twice this amplitude
is a highly nontrivial check of our construction.

The remaining three entries of \tab{DCconstructions} provide
even more constraining checks of our procedure,
as they include the non-supersymmetric chiral-fermion numerators~\eqref{nfermionboth}.
We verified that the double copy,
\be
   {\cal I}_i^{{\cal N}=0+0,{\rm matter}} = \int\!\!\frac{d^{D}\ell}{(2\pi)^{D}} \,
      \frac{n_i^{\rm fermion}\,\overline{n}_i^{\rm fermion}
        \!+ \overline{n}_i^{\rm fermion}\,n_i^{\rm fermion}}{D_i} = 2\,{\cal I}_i^{{\cal N}=0,\rm scalar} \,,
\label{Neq0p0Integrals}
\ee
reproduces twice the scalar-matter amplitude~\cite{Dunbar:1994bn}:
\beal
 & M^{{\cal N}=0,{\rm scalar}}_4(1^{--}\!,2^{--}\!,3^{++}\!,4^{++})
    = \frac{i r_{\Gamma} \braket{12}^4 [34]^4}{ (4\pi)^{D/2} } \frac{1}{2}
      \bigg\{\! - \frac{t^3 u^3}{s^8}
                  \left( \ln^2\!\left( \frac{-t}{-u} \right)\!+ \pi^2 \right) \\ &
                + \frac{(t\!-\!u)(t^4\!+\!9 t^3 u\!+\!46 t^2 u^2\!+\!9 t u^3\!+\!u^4)}
                       {30s^7}
                  \ln\!\left( \frac{-t}{-u} \right) \!
                + \frac{2t^4\!+\!23 t^3 u\!+\!222 t^2 u^2\!+\!23 t u^3\!+\!2u^4}
                       {180s^6}
      \bigg\} \,.
\label{N0gravityscalar}
\eeal
Replacing the left-copy fermion numerators by the chiral ${\cal N}=1$ numerators,
\be
   {\cal I}_i^{{\cal N}=1+0,{\rm matter}} = \int\!\!\frac{d^{D}\ell}{(2\pi)^{D}} \,
      \frac{n_i^{{\cal N}=1,{\rm fund}}\,\overline{n}_i^{\rm fermion}
        \!+ \overline{n}_i^{{\cal N}=1,{\rm fund}}\,n_i^{\rm fermion}}{D_i}= {\cal I}_i^{{\cal N}=2,{\rm matter}}  \,,
\label{Neq1p0Integrals}
\ee
also reproduces the correct amplitude~\eqref{N2gravitymatter}.
Doing the same for ${\cal N}=2$ numerators,
\be
   {\cal I}_i^{{\cal N}=2+0,{\rm matter}} = \int\!\!\frac{d^{D}\ell}{(2\pi)^{D}} \,
      \frac{n_i^{{\cal N}=2,{\rm fund}}\,\overline{n}_i^{\rm fermion}
        \!+ \overline{n}_i^{{\cal N}=2,{\rm fund}}\,n_i^{\rm fermion}}{D_i}= {\cal I}_i^{{\cal N}=2,{\rm vector}}  \,,
\label{Neq2p0Integrals}
\ee
results in the correct ${\cal V}_{{\cal N}=2}$-vector amplitude, which can be otherwise calculated as
\be
        M^{{\cal N}=2,\rm vector}_4(1,2,3,4) =  M^{{\cal N}=4,\rm matter}_4(1,2,3,4) -2M^{{\cal N}=2,\rm matter}_4(1,2,3,4) \,.
\label{N2gravityvector}
\ee

In addition, even if they are not included in \tab{DCconstructions},
we note that the fundamental scalar numerators of \sect{N0sc}
also produce sensible double copies at one loop,
such as
\be
   {\cal I}_i^{{\cal N}=0'+0', {\rm \, scalar \, matter}} =
      \int\!\!\frac{d^{D}\ell}{(2\pi)^{D}} \,
      \frac{n_i^{\rm scalar} \overline{n}_i^{\rm scalar}
        \!+ \overline{n}_i^{\rm scalar} n_i^{\rm scalar}}{D_i}
      = 2\,{\cal I}_i^{{\cal N}=0,\rm scalar} \,,
\label{Neq0p0IntegralsScalar}
\ee
which integrates to twice the scalar-matter amplitude~\eqref{N0gravityscalar}.
However, this is equivalent to the adjoint construction
of refs.~\cite{Bern:2013yya,Nohle:2013bfa}.
\Eqn{N1even} also implies that the double copy
\be
   {\cal I}_i^{{\cal N}=1+0', {\rm matter}} = \int\!\!\frac{d^{D}\ell}{(2\pi)^{D}} \,
      \frac{n_i^{{\cal N}=1,{\rm fund}}\,\overline{n}_i^{\rm scalar}
        \!+ \overline{n}_i^{{\cal N}=1,{\rm fund}}\,n_i^{\rm scalar}}{D_i}= {\cal I}_i^{{\cal N}=2,\rm matter} \,,
\label{Neq1p0IntegralsScalar}
\ee
trivially follows from the adjoint double copy 
$({\cal N}\!=\!2,{\rm adj})\times({\cal N}\!=\!0,{\rm scalar})$,
which integrates to the ${\cal N}=2$ matter amplitude~\eqref{N2gravitymatter}.

While the double copies involving scalars work flawlessly for one-loop amplitudes with external graviton multiplets, we expect similar constructions to be more problematic or ambiguous at higher loops. In particular, as far as we know, the scalars cannot be used to obtain pure gravity amplitudes at two loops and higher. Furthermore, in the absence of supersymmetry, the scalar self-interactions are not fixed by universal considerations.

\subsection{Pure gravity amplitudes}
\label{puregravity}

\begin{figure}[t]
      \centering
      \includegraphics[scale=0.75]{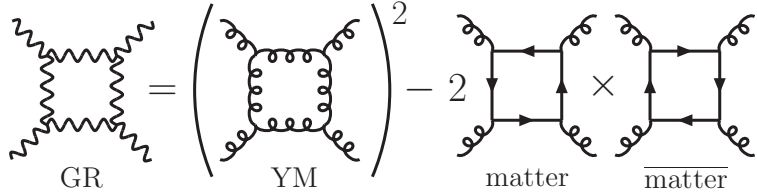}
      \vspace{-5pt}
\caption[a]{\small Construction of a pure gravity amplitude by subtraction of axion and dilaton. At one loop, the matter contribution can be from either a fermion or a scalar.}
\label{SubtractionFigure}
\end{figure}

In the previous section, we checked that our fundamental numerators produce all the matter-amplitude double copies from \tab{DCconstructions}.
Based on that and the conventional adjoint-representation double copies,
we can now formulate what the recipe in \sect{FundDoubleCopy} gives
for pure (super-)gravity four-point\footnote{In principle, the formulas in this subsection are valid for any multiplicity at one loop, once the corresponding numerators are known.} one-loop amplitudes. We will not write out the explicitly integrated forms of the one-loop four-point amplitudes,
as they can be found in ref.~\cite{Dunbar:1994bn}.
After integration, all our amplitudes agree with the results therein.

As in illustrated in \fig{SubtractionFigure},
the integrals for pure Einstein gravity, or general relativity (GR),
in our framework are given by
\be
   {\cal I}_i^{\rm GR} = \int\!\!\frac{d^{D}\ell}{(2\pi)^{D}}
      \frac{(n_i^{\rm YM})^2 - 2\,n_i^{\rm fermion}\,\overline{n}_i^{\rm fermion}}
           {D_i} \,.
\label{Neq0fermIntegrals}
\ee
The pure YM numerators
can be easily constructed from the numerators~\eqref{N4Box},
\eqref{N1evenBox} and \eqref{N0scalarBox} through the well-known~\cite{Bern:1993mq}
one-loop supersymmetry decomposition. This decomposition, for both numerators and amplitudes at one loop, is given by
\beal
   n_i^{\rm YM}& = n_i^{{\cal N}=4} - 4\,n_i^{{\cal N}=2,{\rm adj}}
                                   + 2\,n_i^{\rm adj\,scalar} \,, \\
   {\cal A}^{\rm YM}_m &={\cal A}^{{\cal N}=4}_m - 4\,{\cal A}^{{\cal N}=2,{\rm adj}}_m
                                   + 2\,{\cal A}^{\rm adj\,scalar}_m \,.
\eeal
The fermion ghosts in \eqn{Neq0fermIntegrals}
subtract the dilaton and the axion contributions.
However, due to the fact that at one loop double copies of anti-aligned chiral fermions are equal to scalar double copies, the same integrals can be reproduced simply as
\be
   {\cal I}_i^{\rm GR} = \int\!\!\frac{d^{D}\ell}{(2\pi)^{D}}
                  \frac{(n_i^{\rm YM})^2 -2(n_i^{\rm adj\,scalar})^2}{D_i} \,.
\label{Neq0scIntegrals}
\ee
Although the latter construction is nice,
the higher-loop cut checks from \sect{cutchecks} suggest
that this equality is accidental.
The fundamental-fermion framework resulting in \eqn{Neq0fermIntegrals}
is the one that should be generally valid at higher loops.

The pure ${\cal N}=1$ supergravity amplitude is generated by the following integrals:
\be
   {\cal I}_i^{{\cal N}=1\,{\rm SG}} = \int\!\!\frac{d^{D}\ell}{(2\pi)^{D}}
      \frac{ n_i^{{\cal N}=1\,{\rm  SYM}}\,n_i^{\rm YM}
           - n_i^{{\cal N}=1, {\rm fund}}\,\overline{n}_i^{\rm fermion}
           - \overline{n}_i^{{\cal N}=1, {\rm fund}}\,n_i^{\rm fermion}}{D_i} \,,
\label{Neq1fermIntegrals}
\ee
and the integrals for pure ${\cal N}=2$ supergravity are given by
\be
   {\cal I}_i^{{\cal N}=2\,{\rm SG}} = \int\!\!\frac{d^{D}\ell}{(2\pi)^{D}}
      \frac{ (n_i^{{\cal N}=1\,{\rm  SYM}})^2
           - 2\,n_i^{{\cal N}=1, {\rm fund}} \,
             \overline{n}_i^{{\cal N}=1, {\rm fund}} }{D_i} \,.
\label{Neq2Integrals}
\ee
The numerators for ${\cal N}=1$ SYM can be constructed from the numerators~\eqref{N4Box} and \eqref{N1evenBox} through the following one-loop supersymmetry decomposition~\cite{Bern:1993mq}:
\be
   n_i^{{\cal N}=1\, {\rm  SYM}} = n_i^{{\cal N}=4} - 3\,n_i^{{\cal N}=2,{\rm adj}} \,.
\ee
In \eqns{Neq1fermIntegrals}{Neq2Integrals}, the matter ghosts subtract the contributions
from one and two ${\cal N}=2$ matter multiplets, respectively. 

Alternatively, we can write the pure ${\cal N}=2$ supergravity using the integrals
\be
   {\cal I}_i^{{\cal N}=2\,{\rm SG}} = \int\!\!\frac{d^{D}\ell}{(2\pi)^{D}}
      \frac{ n_i^{{\cal N}=2\,{\rm  SYM}}\,n_i^{\rm YM}
           - n_i^{{\cal N}=2, {\rm fund}}\,\overline{n}_i^{\rm fermion}
           - \overline{n}_i^{{\cal N}=2, {\rm fund}}\,n_i^{\rm fermion}}{D_i} \,,
\label{Neq2fermIntegrals}
\ee
where the ghosts cancel a ${\cal V}_{{\cal N}=2}$ multiplet,
i.e. an abelian vector multiplet. 
The numerators for ${\cal N}=2$ SYM can be constructed
from the numerators~\eqref{N4Box} and~\eqref{N1evenBox}
through the following decomposition~\cite{Bern:1993mq}:
\be
   n_i^{{\cal N}=2\, {\rm  SYM}} = n_i^{{\cal N}=4} - 2\,n_i^{{\cal N}=2,{\rm adj}} \,.
\ee
The fact that the above two very different constructions of pure ${\cal N}=2$ supergravity give the same amplitude is highly nontrivial.
If our prescription does not encounter unexpected obstructions at higher loops,
this nontrivial equality should be true in all generality.

The pure ${\cal N}=3$ supergravity amplitude is obtained through the integrals
\be
   {\cal I}_i^{{\cal N}=3\,{\rm SG}} = \int\!\!\frac{d^{D}\ell}{(2\pi)^{D}}
      \frac{ n_i^{{\cal N}=2\,{\rm  SYM}}\,n_i^{{\cal N}=1\,{\rm  SYM}}
           - n_i^{{\cal N}=2, {\rm fund}}\,\overline{n}_i^{{\cal N}=1, {\rm fund}}
           - \overline{n}_i^{{\cal N}=2, {\rm fund}}\,n_i^{{\cal N}=1, {\rm fund}}}{D_i} \,,
\label{Neq3Integrals}
\ee
where the ghosts subtract out a ${\cal V}_{{\cal N}=4}$-vector multiplet,
equal to the combination of two ${\cal N}=3$ chiral multiplets.

Although pure ${\cal N}=4$ supergravity is a factorizable theory, meaning that its amplitudes and integrals can be written as an adjoint double copy of pure YM theories, 
\be
   {\cal I}_i^{{\cal N}=4\,{\rm SG}} = \int\!\!\frac{d^{D}\ell}{(2\pi)^{D}}
      \frac{ n_i^{{\cal N}=4\,{\rm  SYM}}\,n_i^{\rm YM}}{D_i} \,,
\label{Neq4Integrals}
\ee
they can also be obtained through a fundamental-representation double copy 
\be
   {\cal I}_i^{{\cal N}=4\,{\rm SG}} = \int\!\!\frac{d^{D}\ell}{(2\pi)^{D}}
      \frac{ n_i^{{\cal N}=2\,{\rm  SYM}}\,n_i^{{\cal N}=2\,{\rm  SYM}}
           - 2n_i^{{\cal N}=2, {\rm fund}}\,\overline{n}_i^{{\cal N}=2, {\rm fund}}}{D_i} \,.
\label{Neq4IntegralsFund}
\ee
Similarly to the ${\cal N}=2$ case, that these two very different constructions
of pure ${\cal N}=4$ supergravity give the same amplitude
is a nontrivial result~\cite{Tourkine:2012vx,Carrasco:2012ca}.

With that said, we have gone through all the cases listed in \tab{DCconstructions}.
Our double-copy prescription produces the correct one-loop four-point amplitudes
for ${\cal N}=0,1,2,3,4$ pure supergravities,
as confirmed by comparing to the previously known results~\cite{Dunbar:1994bn}.

\section{Conclusion and outlook}

In this paper, we have extended the scope of color-kinematics duality to 
matter fields in the fundamental representation.
As we showed through various examples, this allows us to construct gravity scattering amplitudes in a broad range of (super-)gravity theories using the double-copy prescription. This includes Einstein gravity, pure ${\cal N}<4$ supergravity, and supergravities with arbitrary non-self-interacting matter. 

The main focus of this work is to address the issue
of unwanted matter degrees of freedom propagating in the loops
that appear when one attempts to construct pure ${\cal N}<4$ supergravities
using the color-kinematics duality.
Such supersymmetric dilaton-axion matter is inherent
to the standard double copy of adjoint-representation gauge theories.
Using the fundamental representation for YM theory matter gives us a means to differentiate these fields from the adjoint ones, so that the double copies can be performed separately.
If we promote the matter double copies to be ghosts,
they cancel the unwanted matter states in the double copy of the vector fields.
Moreover, the ghost double-copy prescription,
$c_i \rightarrow (-1)^{|i|} \overline{n}'_i$,
can be easily extended to a tunable-matter prescription,
$c_i \rightarrow (N_X)^{|i|} \overline{n}'_i$,
thus producing theories with any number of matter multiplets coupled to gravity.

We have presented nontrivial evidence supporting our framework
using examples both for tree and loop-level amplitudes, as well as more general arguments.
At tree level, we have checked the construction of gravitational amplitudes
with non-self-interacting matter though seven points, with the most general external states.
At one loop, we have constructed the color-kinematics representation
of four-gluon amplitudes with general fundamental matter circulating in the loop.
Using the double-copy prescription, we have obtained the one-loop four-graviton amplitudes
in Einstein gravity, pure ${\cal N}<4$ supergravity and supergravity with generic matter.

Our explicit calculations give novel and simple forms for YM numerators containing ${\cal N}=2,1$ supersymmetric fundamental matter, as well as non-supersymmetric fundamental scalars and fermions. Equivalent numerators that satisfy the color-kinematics duality were known~\cite{Carrasco:2012ca,Bern:2013yya,Nohle:2013bfa,Ochirov:2013xba} for all but the odd part of ${\cal N}=1$ the matter contribution, which we give in \eqn{N1oddBox}. After integration, our new representations of one-loop gravity amplitudes are in full accord with the results of Dunbar and Norridge~\cite{Dunbar:1994bn}.
Moreover, they provide a direct check of the nontrivial equality of the symmetric
${\cal N}=1+1$ and the asymmetric ${\cal N}=2+0$ supergravity construction.

At two loops, we have checked the consistency of the construction of  Einstein gravity and pure ${\cal N}<4$ supergravity.  We considered four-dimensional unitarity cuts of two-loop amplitudes, and applied the ghost prescription to the matter double copies. The cuts show that, when using double copies of fundamental fermions, our construction correctly eliminates the unwanted dilaton-axion degrees of freedom naturally present in the double copy of pure YM theory.
We have checked that the same cancellation happens in the supersymmetric generalizations. The two-loop checks also show that when using double copies of fundamental scalars the cancellation with the dilaton-axion states is incomplete. Table \ref{DCconstructions} summarizes the valid double copies for the matter ghosts.

While the double copy of fundamental scalars is unsuitable for the construction of Einstein gravity, it is interesting to note that the fundamental-scalar amplitudes nontrivially satisfy the color-kinematics duality. The resulting double-copy amplitude should correspond to some gravitational amplitudes that are corrected by four-scalar terms and possibly higher-order interactions. We leave the details of the non-supersymmetric double copy of fundamental scalars for future work.

\begin{table*}
\centering
\begin{tabular}{|c|c|c|r|c|}
\hline 
\!\!$D$\!\! &
\!$A^\mu\!\otimes\!A^\nu\!= \phi \oplus\!h^{\mu\nu}\!\!\oplus\!B^{\mu\nu}$\!\! &
tensoring matter &
\!\!$({\rm matter})^2\!= \phi \oplus\!B^{\mu\nu}\!\!\oplus\!D^{\mu \nu\rho\sigma}$\!\! &
$\Rightarrow$ states \\
\hline 
$3$ & 
$1 \otimes 1 = 1 \oplus 0 \oplus 0$ & 
$\lambda \otimes \overline \lambda$\! & 
$1 \otimes 1 = 1 \oplus 0 \oplus 0$\! & 
\!topological \hskip -0.4cm \phantom{\Big[} \\
\hline
$4$ & 
$2 \otimes 2 = 1 \oplus 2 \oplus 1$  & 
\!\!$(\lambda^+\!\!\otimes\!\lambda^-)\!\oplus\!(\lambda^-\!\!\otimes\!\lambda^+)$\!\! &
\!$(1 \otimes 1)\!\oplus\!(1 \otimes 1) = 1 \oplus 1 \oplus 0$\! & 
$h^{\mu\nu}$ \hskip -0.4cm \phantom{\Big[} \\
\hline
$5$ &
$3 \otimes 3 = 1 \oplus 5 \oplus 3$   & 
$\lambda^\alpha\!\otimes \overline \lambda^\beta$\! & 
\!$2 \otimes 2 = 1 \oplus 3 \oplus 0$\! & 
$h^{\mu\nu}$ \hskip -0.4cm \phantom{\Big[} \\
\hline
$6$ &
$4 \otimes 4 = 1 \oplus 9 \oplus 6$  & 
\!\!$(\lambda^\alpha\!\otimes\!\lambda^\beta)\!\oplus\!
     (\widetilde \lambda^{\dot \alpha}\!\otimes\!\widetilde \lambda^{\dot \beta})$\!\! & 
\!$(2 \otimes 2)\!\oplus\!(2 \otimes 2) = 1 \oplus 6 \oplus 1$\! & 
\!$h^{\mu\nu},D^{\mu \nu\rho\sigma}_{\rm ghost}$ \hskip -0.4cm \phantom{\Big[} \\
\hline
\!\!\!$10$\!\!\!&
$8 \otimes 8 = 1 \oplus 35 \oplus 28$  & 
$\lambda^A\!\otimes \lambda^B$\! & 
$8 \otimes 8 = 1\,\oplus \,28 \,\oplus \,35$\! & 
\!$h^{\mu\nu}, D^{\mu \nu\rho\sigma}_{\rm ghost}$ \hskip -0.4cm \phantom{\Big[} \\
\hline
\end{tabular}
\caption[a]{\small On-shell state counting for pure bosonic gravities in various spacetime dimensions. Subtracting the matter double copies from the vector double copies gives the correct states in the pure theories in dimensions $D=3,4,5$, but for $D=6,10$ the naive approach gives excess cancellations resulting in a four-form ghost. }
\label{DdimConstructions}
\end{table*}

While our construction of pure supergravities passes many nontrivial checks, it is well known that higher-loop calculations can be plagued by subtleties coming from dimensional regularization. Thus it is important that our prescription is carefully scrutinized by explicit $L>1$ calculations in both pure and matter-coupled supergravities. Starting at two loops, the scheme-dependence of different types of dimensional-regularization prescriptions would be interesting to study.

It would be also interesting to see how our double-copy prescription for pure gravities
can be extended beyond $D=4-2\epsilon$ dimensions.
For instance, in \tab{DdimConstructions}, we provide some
prospective integer-dimensional constructions of this kind,
ranging from $D=3$ to $D=10$ in the non-supersymmetric setting. As shown, starting in $D=6$ the naive ghost prescription appears to need a modification in order to avoid over-cancellation of the physical states.

In conclusion, we expect that our results will open a new window into the study of pure ${\cal N}=0,1,2,3$ supergravity multiloop amplitudes and their ultraviolet properties.
Pure Einstein gravity is known to have a divergence in the two-loop effective action, as was proven in refs.~\cite{Goroff:1985sz,Goroff:1985th,vandeVen:1991gw}
using computerized symbolic manipulations. The ${\cal N}=1,2,3$ theories are known to have no divergences at one and two loops due to supersymmetry~\cite{Grisaru:1976nn,Tomboulis:1977wd,Deser:1977nt,Howe:1980th},
but the status of these theories at three loops and beyond is unknown.
It would be interesting to reinvestigate the ${\cal N}<4$ theories from the modern point of view using new analytic methods and structures, including the current results, in hope of gaining better understanding of the structure of quantum gravity.

\section*{Acknowledgments}
\vskip -.25 cm
We would like to thank Zvi Bern, Ruth Britto and John Joseph Carrasco for helpful discussions. We also thank Marco Chiodaroli, Murat G\"{u}naydin, Radu Roiban and Piotr Tourkine for conversations and collaborations on topics related to this work.


\appendix

\section{Specific matter interactions}
\label{AppendixA}

In this appendix, we list the types of YM interactions that we consider in this paper. First of all, let us separate the pure vector interactions from the matter contributions, then the Lagrangian is
\be
{\cal L}={\cal L}_{\rm pure\,SYM}+{\cal L}_{\rm matter} \,.
\ee
For example, in the non-supersymmetric case
$ {\cal L}_{\rm pure\,YM} = -\frac{1}{4} F^a_{\mu\nu} F^{a\mu\nu} $, with the following normalization for the field strength
$ F^a_{\mu\nu} = \partial_{[\mu} A^a_{\nu]}
               - ig/\sqrt{2} \tf^{abc}A^{b}_{\mu}A^{c}_{\nu} $. 

In case we have a single fundamental complex scalar, the matter Lagrangian is 
\be
   {\cal L}_{\rm scalar} = (\overline{D_\mu \phi})_{\bar \jmath} (D^\mu \phi)_j
                 - \frac{g^2}{4} (\overline{\phi}_{\bar i} T^a_{i \bar \jmath} \phi_j)
                                 (\overline{\phi}_{\bar k} T^a_{k \bar l} \phi_l) \,,
\ee
where
$(D_\mu \phi)_j = \partial_\mu \phi_j -ig/\sqrt{2}\,T^{a}_{j \bar k} A_\mu^{a} \phi_k$. Note that we include a four-scalar contact term
since it is needed when generalizing to supersymmetric theories below.
We still regard theories with such a contact term ``non-self-interacting''
in the sense that they contain no self-interactions
beyond those required by supersymmetry and gauge invariance.

For a single fundamental Weyl fermion $\psi_k$, the matter term is the usual one
\be
{\cal L}_{\rm fermion}= \widetilde{\psi}_{\dot \alpha \bar \jmath}\,
      i \s D^{\dot \alpha \alpha}_{j \bar k} \psi_{\alpha k} \,,
\ee
with the covariant derivative
$ \s D^{\dot \alpha \alpha}_{j \bar k}
  = \overline{\sigma}^{\dot \alpha \alpha}_{\mu}
    \big( \delta_{j \bar k} \, \partial^\mu - i g/\sqrt{2} A^\mu_{j \bar k} \big) $.
In our convention, the field $\psi$ ($\widetilde{\psi}$) carry the on-shell mode corresponding to a negative (positive) helicity fermion. 
    
For the supersymmetric cases, the kinetic parts of the matter interactions are similar when written in terms of the individual physical components of the multiplet. Additionally, due to supersymmetry, Yukawa interactions between the adjoint fermion $\lambda$ (and scalar $\varphi_{12}$) in the vector multiplet and the matter fields are also present.
Explicitly, the matter contributions are
\be
   {\cal L}_{{\cal N}=1\,{\rm matter}} =
        {\cal L}_{\rm scalar}
      + \widetilde{\psi}_{\dot \alpha \bar \jmath} \,
        i \s D_{j \bar k}^{\dot \alpha \alpha} \psi_{\alpha k}
      + g \widetilde{\psi}_{\dot \alpha \bar \jmath}
          \widetilde{\lambda}_{j \bar k}^{\dot \alpha} \phi_k
      + g \overline{\phi}_{\bar \jmath}
          \lambda_{j \bar k}^{\alpha} \psi_{\alpha k} \,,
\ee
and
\be
   {\cal L}_{{\cal N}=2\,{\rm matter}} =
      \widetilde{\psi}_{\dot \alpha \bar \jmath} \,
      i \s D_{j \bar k}^{\dot \alpha \alpha} \psi_{\alpha k}
    + \psi^{\alpha}_{\bar \jmath} \,
      i \s D_{\alpha \dot \alpha j \bar k} \widetilde{\psi}_{k}^{\dot \alpha}
    + g{\cal L}^{\rm Yukawa}_{\rm matter}
    + {\cal L}^{\rm bosonic}_{\rm matter} \,,
\ee
where the Yukawa interactions are
\be
   {\cal L}^{ \rm Yukawa}_{\rm matter} =
      ( \widetilde{\psi}_{\dot \alpha \bar \jmath}
        \widetilde{\lambda}^{\dot \alpha A}_{j \bar k}
      + \psi_{\bar \jmath }^\alpha \lambda^A_{\alpha j \bar k} ) \phi_{A k}
    + \overline{\phi}^A_{\bar \jmath}
      ( \lambda^{\alpha}_{A j \bar k} \psi_{\alpha k}
      + \widetilde{\lambda}_{\dot \alpha A j \bar k}
        \widetilde{\psi}_{k}^{\dot \alpha} )
    + \psi_{\bar \jmath}^\alpha \varphi_{12 j \bar k} \psi_{\alpha k}
    + \widetilde{\psi}_{\dot \alpha \bar \jmath}
      \overline{\varphi}^{12}_{j \bar k}
      \widetilde{\psi}_{k}^{\dot \alpha} \,,
\ee
and the purely-bosonic matter is given by
\beal
   {\cal L}^{\rm bosonic}_{\rm matter} = 
      (\overline{D^\mu\phi^A})_{\bar \jmath} (D_\mu\phi_A)_{j}
    - \frac{g^2}{2}\, \overline{\phi}_{\bar \jmath}^A
      \{ \overline{\varphi}^{12}, \varphi_{12} \}_{j \bar k}\, \phi_{A j}
    + \frac{g^2}{4} \big( \overline{\phi}^A_{\bar \imath}
                          T^{a}_{i \bar \jmath} \phi_{A j} \big)
                    \big( \overline{\phi}^B_{\bar k}
                          T^{a}_{k \bar l} \phi_{B l} \big) & \\
    \null - \frac{g^2}{2} \big( \overline{\phi}^A_{\bar \imath}
                          T^{a}_{i \bar \jmath} \phi_{B j} \big)
                    \big( \overline{\phi}^B_{\bar k}
                          T^{a}_{k \bar l} \phi_{A l} \big) & \,.
\eeal

\bibliographystyle{JHEP}
\bibliography{references}

\end{document}